\documentclass[11pt,draftclsnofoot,onecolumn]{IEEEtran}
\usepackage{cite}
\usepackage{amsmath,amssymb,amsfonts}
\usepackage{algorithmic}
\usepackage{graphicx}
\usepackage{textcomp}
\usepackage{booktabs}
\usepackage{epstopdf}
\usepackage{array}
\usepackage{pgffor}
\usepackage{bm}
\usepackage{subfigure}
\usepackage{color}
\usepackage{setspace}
\usepackage{tcolorbox}
\usepackage{multicol}
\usepackage{capt-of}
\usepackage{acronym}

\acrodef{ista}[ISTA]{iterative shrinkage thresholding algorithm}	
\acrodef{lista}[LISTA]{learned \ac{ista}}	
\acrodef{em}[EM]{electromagnetic}
\acrodef{doi}[DoI]{domain of investigation}
\newcommand{\Revise}[1]{{\color{black}{#1}}}

\begin{document}
	\title{Physics Embedded Machine Learning for Electromagnetic Data Imaging
	}
	
	\author{Rui Guo,  \IEEEmembership{Member, IEEE}, Tianyao Huang, \IEEEmembership{Member, IEEE},  Maokun Li, \IEEEmembership{Senior Member, IEEE} , Haiyang Zhang, \IEEEmembership{Member, IEEE}, and Yonina C. Eldar, \IEEEmembership{Fellow, IEEE}
		
		\vspace{-1.85cm}
		\thanks{R. Guo, T. Huang, M. Li are with Department of EE, Tsinghua University, Beijing, China. (e-mail: \{maokunli, guor93, huangtianyao\}@tsinghua.edu.cn). H. Zhang and Y. C. Eldar are with the Faculty of Math and CS, Weizmann Institute of Science, Rehovot, Israel (e-mail: \{haiyang.zhang, yonina.eldar\}@weizmann.ac.il). This work is supported in part by the Institute for Precision Medicine, Tsinghua University, National Natural Science Foundation of China (61971263 and 62171259), and Shuimu Tsinghua Scholar Program of Tsinghua University (2021SM031).  
		}
		%
	}
	\maketitle
	\begin{abstract}
		Electromagnetic (EM) imaging is widely applied in sensing for security, biomedicine, geophysics, and various industries. It is an ill-posed inverse problem whose solution is usually computationally expensive. Machine learning (ML) techniques and especially deep learning (DL) show potential in fast and accurate imaging. However, the high performance of purely data-driven approaches relies on constructing a training set that is statistically consistent with practical scenarios, which is often not possible in EM imaging tasks. Consequently, generalizability becomes a major concern. On the other hand, physical principles underlie EM phenomena and provide baselines for current imaging techniques. To benefit from prior knowledge in big data and the theoretical constraint of physical laws, physics embedded ML methods for EM imaging have become the focus of a large body of recent work. 
		
		This article surveys various schemes to incorporate physics in learning-based EM imaging. We first introduce background on EM imaging and basic formulations of the inverse problem. We then focus on three types of strategies combining physics and ML for linear and nonlinear imaging and discuss their advantages and limitations. Finally, we conclude with open challenges and possible ways forward in this fast-developing field. Our aim is to facilitate the study of intelligent EM imaging methods that will be efficient, interpretable and controllable. 
		
	\end{abstract}
	\vspace{-0.6cm}
	\section{Introduction}
	
	Electromagnetic (EM) fields and waves have long been used as a sensing method. This is because the EM field can penetrate various media,  interact with materials, and alter its distribution in both space and time. Hence electric and magnetic properties of materials, such as {\color{black}permittivity, permeability and conductivity}, can be inferred from samples of the field. EM imaging refers to reconstructing the value distribution of electric or magnetic parameters from measured EM fields, through which a better understanding of the domain of investigation (DoI) can be obtained. EM imaging techniques have been widely applied in security, biomedicine, geophysics, and various industries. In security, EM imaging for example using radars can help locate targets that are invisible to optical imaging. In biomedicine, microwave imaging can detect anomalies in the permittivity distribution caused e.g. by cerebral hemorrhage. As a final example, images of conductivity distribution reconstructed from low-frequency EM fields may reveal deep structures in the earth. 
	
	\begin{figure}[!]
		\centering
		\includegraphics[width=160mm]{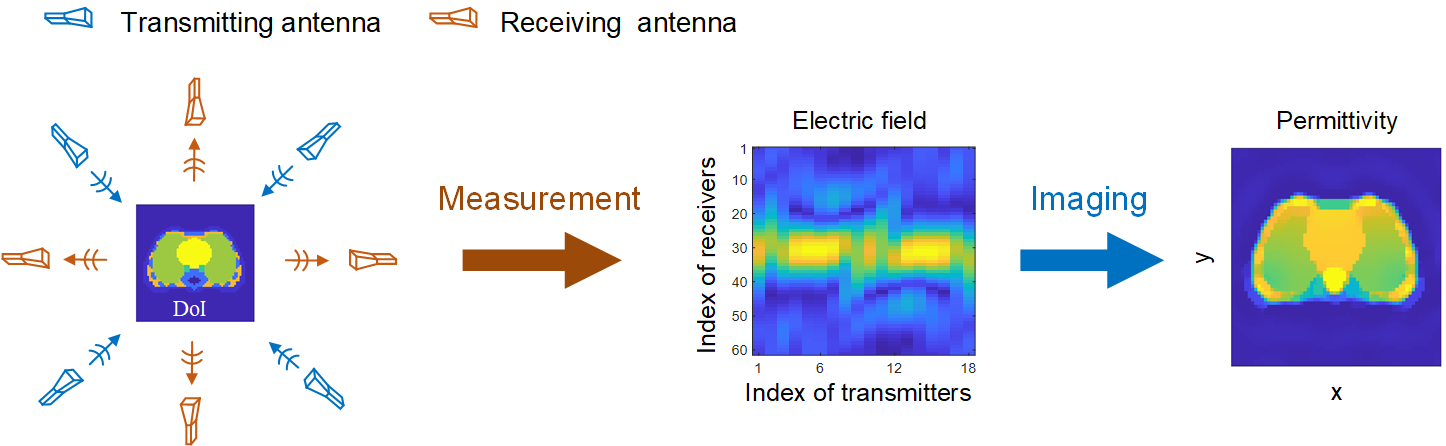}	
		\caption{The setup of EM imaging. EM imaging converts the measured data to the spatial distribution of electric parameters in the \ac{doi}.    }
		\label{fig1}
	\end{figure}
	
	A theoretical model of EM imaging is illustrated in Fig.~\ref{fig1}, where EM sensors, i.e. antennas, are deployed around the \ac{doi}. When an external source illuminates the \ac{doi}, the sensors record the EM field. Given the transmitting waveform as well as the locations of transmitting antennas, the EM field propagates according to Maxwell's equations \cite{chew1990waves}. In the frequency domain, {\color{black}EM propagation can be described by the following partial differential equation,} \vspace{-0.3cm} 
	\begin{equation}\label{eq1}
		\nabla \times \nabla\times \mathbf E(\mathbf r) - \omega^2\mu\epsilon(\mathbf r)\mathbf E(\mathbf r) 
		= i\omega\mu \mathbf J (\mathbf r),\vspace{-0.3cm} 
	\end{equation}
	where $\mathbf E$ is the vector electric field, $\mathbf r$ is the spatial position, $\mu$ is permeability, $ \epsilon $ is complex permittivity, $ \mathbf J $ is the electric current source, $ \omega $ is the angular frequency, and $ \nabla\times $ is the curl operator. {\color{black}The complex permittivity is expressed as $ \epsilon = \epsilon_R + i\sigma/\omega $, where its real part $ \epsilon_R $ is permittivity and its imaginary part is related to conductivity $ \sigma $. Equation (\ref{eq1}) can describe both wave physics ($ \epsilon_R \gg  \sigma/\omega $) and diffusion physics ($ \sigma/\omega \gg \epsilon_R$), depending on the settings of investigation.} Here permeability is assumed to be constant, which is reasonable in most imaging scenarios. In EM imaging, we usually have information about the sources; therefore $\omega$ and $\mathbf J$ are both known. Once we measure the electric field at the receiver locations, the {\color{black}complex} permittivity can be {\color{black}recovered} by solving the above equation. This process defines the EM inverse problems, in which the electromagnetic parameters are solved given measured electric fields. 
	
	The EM inverse problem is nonlinear and often ill-posed due to several challenges:
	\begin{itemize}
		\item The nonlinearity comes from complex interactions between the measured EM field and the material parameters. As we can see from (\ref{eq1}), the product of $\mathbf E$ and $\epsilon$ results in a nonlinear relationship where the nonlinearity increases with $\epsilon$.
		
		\item {\color{black}The ill-posedness arises from multiple scatterings, insufficient measurements, and noise corruption. Due to multiple scatterings of EM waves, a slight variation of targets may change the EM field substantially}. In most cases, we only record the field at specific locations, i.e., the field samples are sparse in space. 
		{\color{black}The attenuation of EM fields caused by diffraction and absorption of media, as well as a noisy environment, further increases the ill-posedness. }
		
		\item Solving EM imaging problems often requires accurate modeling of EM wave propagation in the \ac{doi}. This process is called forward modeling and is implemented by numerical algorithms such as the finite element method. However, it is computationally intensive, especially for large \ac{doi}s.
	\end{itemize}

	{\color{black}Recent advances in big data storage, massive parallelization, and optimization algorithms facilitated the development of machine learning (ML) and its applications in EM imaging \cite{massa2019dnns,wang2021deep,li2021machine,chen2020review}. 
		ML is attractive for overcoming the above limitations due to the following aspects. First, the time-consuming operations of modeling and inversion can be surrogated by data-driven models to make imaging faster. Second, prior knowledge that is difficult to describe with rigorous forms can be recorded after the learning process, which helps improve imaging accuracy. Finally, DL software frameworks provide user-friendly interfaces to fully exploit the computing power without low-level programming on heterogeneous platforms, which largely reduces the complexity of algorithm implementation for high-performance imaging.
		
		Training a surrogate model for data-image mappings such as deep neural networks (DNNs) has shown promising results \cite{wang2021deep}. However, the success relies on constructing a training dataset that is statistically consistent with practical scenarios. {\color{black}Due to the multiple scattering effects, simply establishing the mapping from EM data to electric properties by ``black-box'' regression may lead to implausible predictions even when the measured data is not highly out-of-distribution. On the other hand, physical laws provide baselines for EM imaging. The relationship between EM fields and electric properties is inherent in Maxwell's equations.}
		
		{\color{black}Recent trends show that a hybrid of physics- and data-driven methods  
			can analyze and predict data more effectively \cite{chen2020review}. Such methods can be grouped into  learning-assisted physics-driven approaches and physics embedded ML approaches. The first category solves the inverse problem in physics-based frameworks, where learning approaches are applied to augment the performance, such as generating better initial guesses \cite{sanghvi2019embedding}, improving the bandwidth of measured data \cite{lin2021low}, or encoding prior knowledge \cite{bora2017compressed}. The second category performs imaging mainly in data-driven manners, where algorithms are designed according to physical laws, such as tailoring inputs and labels \cite{8565987,wei2018deep,ye2020inhomogeneous,9751403}, loss functions \cite{jin2020physics,raissi2019physics,bar2021strong}, and neural network structures \cite{9539099,8434321,guo2021physics1}. While frameworks of physics-based techniques have been well studied, learning methods for EM imaging vary widely.}
		
		This article aims to review recent frontiers in physics embedded ML for EM imaging techniques, and shed insight on designing efficient and interpretable ML-based imaging algorithms. Existing approaches include modeling Maxwell's equations into the learning process and combining trainable parameters with full-wave EM solvers or differential/integral operators \cite{shahriari2021error,hu2021theory,bar2021strong,guo2020pixel,guo2021physics1,guo2021physics2,Fu2021Toep,fu2022block,jin2020physics,shan2021neural,9539099,8434321}. These ML models not only describe physical principles but also record the prior knowledge gained from the training data.  Compared with purely data-driven models, the physics embedded approaches possess higher generalizability and can learn effectively from less training data \cite{shlezinger2020model}. In addition, since there is no need to train the physics part, both the memory and computation complexity of DNNs can be reduced. 
		
		We begin in Sec.~\ref{sec:formulation} by introducing basic formulations of the EM imaging problem and stating some of the challenges with conventional methods. We then categorize existing physics embedded ML approaches into three kinds: \textit{learning after physics processing}, \textit{learning with physics loss}, and \textit{learning with physics models}, presented in Sec.~\ref{sec:sequential} to \ref{sec:unroll}, respectively. {\color{black}Sec.~\ref{sec:discussion}  discusses open challenges and opportunities in this fast-developing field.  We draw conclusions in Sec.~\ref{sec:conclusion}.}
		
	}
	\vspace{-0.2cm}
	\section{Formulations and challenges of EM imaging}
	\label{sec:formulation}
	EM imaging is an inverse problem that calculates the electric parameters of the \ac{doi} from measured EM fields. This process incorporates EM modeling that simulates the ``measured'' data based on numerical models. It can be described as minimizing the misfit between the observed and simulated data \cite{habashy2004general},\vspace{-0.4cm}
	\begin{equation}\label{eq4}
		L(\boldsymbol{\epsilon}) =   \|\mathbf d_\text{obs} - F(\boldsymbol{\epsilon})\|^2 + \lambda \phi_r(\boldsymbol{\epsilon}), \vspace{-0.4cm}
	\end{equation}
	where $ \mathbf d_\text{obs} $ is the field observed by receivers, $ \boldsymbol{\epsilon} $ is {\color{black}complex permittivity}\footnote{\color{black}Note that  complex permittivity is usually simplified to conductivity in low-frequency EM methods. This article uses complex permittivity to represent the unknown in both high- or low-frequency methods. } , $ F(\boldsymbol{\epsilon}) $ represents the EM  modeling function, $ \phi_r $ is the regularization term and $ \lambda $ is a regularization factor. 
	
	EM modeling, $ F(\boldsymbol{\epsilon}) $, computes the EM field in space given the permittivity distribution in the \ac{doi} and the information on the sources. It is usually achieved by numerical methods, e.g., the finite element method, the finite difference method, and the methods of moments \cite{jin2011theory}. These techniques partition the \ac{doi} into thousands or millions of subdomains and convert the wave equation to a matrix equation. EM fields in space are obtained by solving the matrix equation that involves thousands or millions of unknowns. The solution process can take minutes or hours. This computationally expensive process is usually called full-wave simulation. To accelerate the modeling process, one can make approximations so that the EM field is linear in the electric parameters, such as the Born or Rytov approximations \cite{chew1990waves}. In this linearized process, the EM field is computed by simple matrix operations, such as matrix-vector multiplications or the Fourier transform.

	The regularization, $ \phi_r(\boldsymbol{\epsilon}) $, is used to incorporate prior knowledge into the imaging process, and it varies in different tasks. For instance, in geophysical or biomedical imaging, to emphasize the sharpness or smoothness of material boundaries, $\ell_1$ and $\ell_2$ norms of the spatial gradient of $ \boldsymbol{\epsilon} $ are usually adopted{\color{black}\cite{klose2022laterally,zhdanov2009new,vignoli2015sharp,abubakar2002imaging,zhong2021electrical}}. In radar imaging, the sparsity of the observed scene is often exploited to improve imaging quality by incorporating sparsity regularization, such as the $\ell_1$ norm given by $\phi_r(\boldsymbol{\epsilon}) = \left\| \boldsymbol{\epsilon} \right\|_1$ {\color{black}\cite{vignoli2005focusing,Potter2009_radarlp,Patel2009_SARl1}}.  
	
	Equation (\ref{eq4}) is usually minimized by iterative gradient descent methods, and some challenges still exist. For instance, each iteration requires computing the forward problem and its Fréchet derivative. Most forward problems are computationally intensive,  and computing the Fréchet derivative of $ F(\boldsymbol{\epsilon}) $ with respect to $ \boldsymbol{\epsilon} $, i.e., $ \mathbf S = \partial F(\boldsymbol{\epsilon})/ \partial \boldsymbol{\epsilon}$, needs to call the forward solver $ F(\boldsymbol{\epsilon}) $ many times, which exacerbates the computational burden. {\color{black}Some fast algorithms that compute  approximations of Fréchet derivatives have been proposed \cite{zhdanov20003d,christiansen2016efficient,habashy2004general}, but the solution process still needs to be accelerated.} Furthermore, when $  F(\boldsymbol{\epsilon})  $ is rigorously solved from Maxwell's equations, the objective function is non-convex and has numerous local minima, due to the nonlinearity between EM responses and permittivity. {\color{black}Finally, gradient descent methods lack flexibility in exploiting prior knowledge that is not described by simple regularization}. {\color{black}On the other hand, stochastic inversion schemes are developed to cope with the necessity of reliable uncertainty estimation and to have a natural way to inform imaging with realistic and complex prior information \cite{shen2018data,hansen2021efficient,de2019probabilistic}; however, stochastic sampling is often computationally intensive.}
	
\begin{figure}[!]
		\centering
		\vspace{-0.4cm}
		\includegraphics[width=130mm]{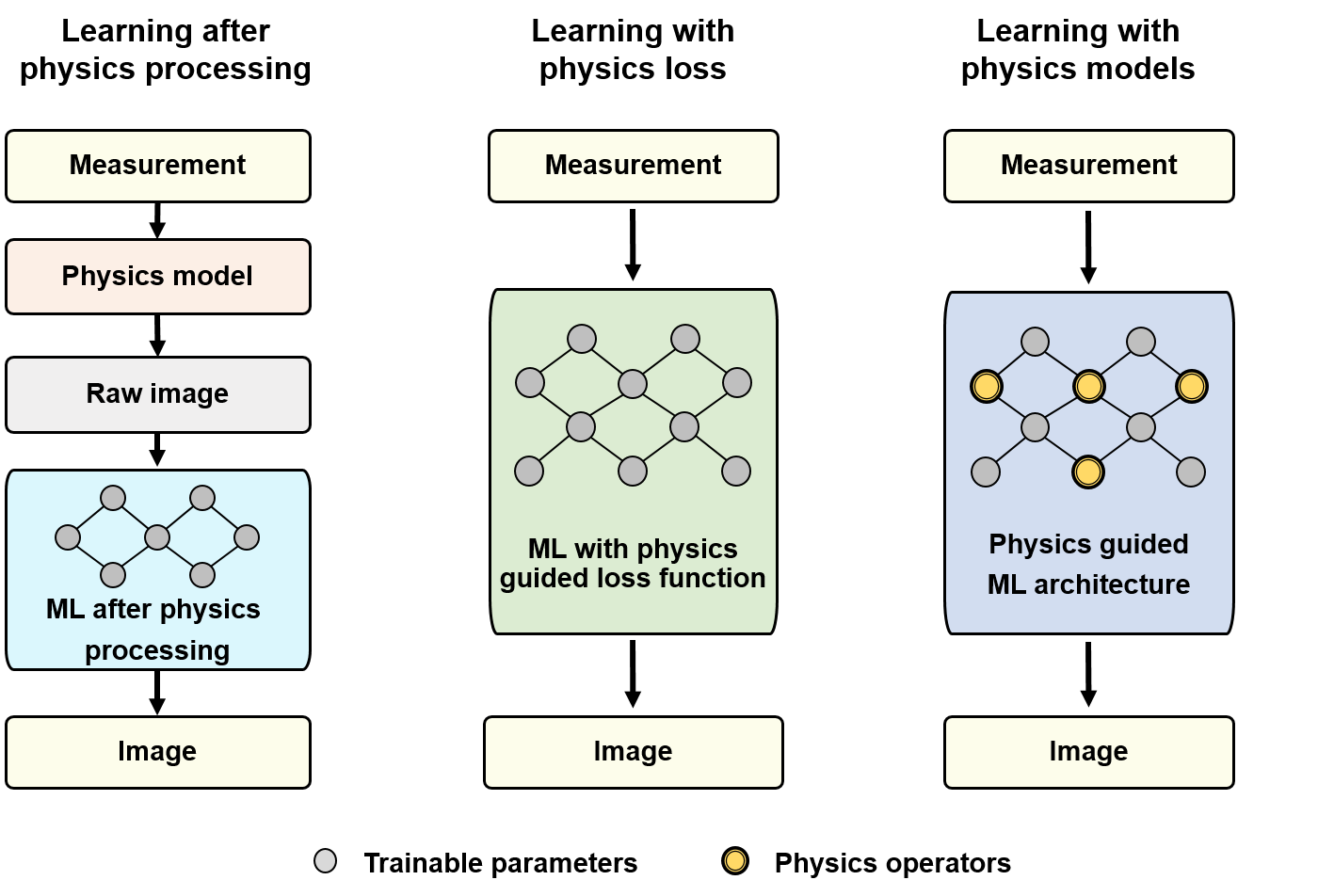}	
		\caption{Three ways of incorporating physics into the ML model. (a) Learning after physics processing: physics model is employed to initialize the input of ML models. (b) Learning with physics loss: physics knowledge is incorporated into the loss functions. (c) Learning with physics models: physics knowledge is used to guide the design of ML architecture.}
		\label{fig3}
	\end{figure}

	Physics embedded ML models provide potential solutions to the challenges mentioned above. In the following, we present three types of physics embedded models for EM imaging, as depicted in Fig.~\ref{fig3}. The first class  processes EM data using conventional physical methods and ML models sequentially \cite{8565987,wei2018deep,ye2020inhomogeneous,xu2020deep,8476623,xiao2019fast}. The second type optimizes network parameters with physics constraints, for example, solving {\color{black}forward problems} in the training loss function \cite{jin2020physics,shahriari2021error,bar2021strong}. The third category unrolls the physical methods with neural networks \cite{Fu2021Toep,fu2022block,guo2020pixel,hu2021theory,guo2021physics2,guo2021physics1,shan2021neural,9539099,8434321}. These three methods will be discussed in detail in the following three sections.
	
	\vspace{-0.2cm}
	
	\section{Learning after physics processing}
	\label{sec:sequential}
	
	The learning after physics processing approach consists of two sequential steps: first, a roughly estimated image is recovered using classical qualitative or quantitative methods; second, the rudimentary image is polished using a DNN trained with the ground truth as labels. In this approach, the tasks of DNNs become image processing, such as eliminating artifacts and improving resolution. We next introduce several classical works that utilize DNNs to enhance image quality in this direction.
	
	In conventional EM imaging, permittivity is iteratively refined by updating in the descent directions of the objective function. This motivates employing multiple convolutional neural network (CNN) modules to progressively improve image resolution starting from some rudimentary images. These images can come from conventional imaging methods,  including linear back-projection \cite{8565987}, subspace optimization method \cite{wei2018deep}, one-step Gauss-Newton method \cite{ye2020inhomogeneous}, and contrast source inversion \cite{xu2020deep}. Finally, the CNN can output super-resolution images close to the ground truth.  To construct the training dataset, many researchers convert the handwritten digits dataset to permittivity models, then perform full-wave simulations to obtain the scattered electric data. To train the DNN, the rudimentary image is taken as the input, while the corresponding true permittivity image is the label. After training with handwritten letters, the DNN predicts targets with more complex shapes and permittivity.
	
	Advanced DNN architectures may improve the performance of image enhancement. The U-Net \cite{chen2020review} is one of the most widely used architectures, which is built on the encoder-decoder architecture and has skip connections bringing encoded features to the decoder. This ensures feature similarity between the input and output and is especially suitable for super-resolution. For example, the authors in \cite{xu2020deep} and \cite{8476623} use U-Nets to achieve super-resolution for 2D microwave imaging. In \cite{xiao2019fast}, the three dimensional (3D) inverse scattering problem is solved by a 3D U-Net, where the input is the preliminary 3D model recovered by Born approximation inversion and the Monte Carlo method.
	
	Another architecture for super-resolution is the generative adversarial network (GAN).  The GAN with cascaded object-attentional super-resolution blocks is applied to imaging with an inhomogeneous background \cite{ye2020inhomogeneous,9751403}. The authors use a GAN with an attention scheme to improve the resolution by highlighting scatterers and inhibiting the artifacts. {\color{black}In \cite{ye2020inhomogeneous}, after training with 6000 handwritten digit scatterers, the GAN reconstructs U-shape plexiglass scatterers in a through-wall imaging test within one second. The structural similarity improves over 50\% compared with conventional algorithms.}

	It should be noted that the more the input is processed by physics, the better the generalizability will be. For instance, \cite{8476623} compares performances with various network inputs, including raw scattered data, permittivity from back-projection, and permittivity from the dominant current scheme (DCS). The network behaves the poorest when directly inputting the raw data, while the best when inputting the image from DCS. This is because the preprocessing in DCS involves more physics and thus reduces the network's burden. A similar conclusion is drawn in \cite{wei2019physics}, where the input and output of neural networks are preprocessed with the wave propagation operator, i.e., Green's function, to a deeper degree, {\color{black}achieving better performance on accuracy and robustness against noise than DCS \cite{8476623}   when recovering high permittivity targets.}
	
	
	The sequential workflow provides great flexibility in borrowing well-developed DL techniques in image processing, and most of the mentioned works can achieve real-time imaging. Recent works have extended this approach to uncertainty quantification of imaging results \cite{wei2020uncertainty}. However, while a DNN can generate a plausible image, the recovered permittivity values may significantly differ from true values. This is because the DNN is trained without the supervision of the EM field. We introduce another group of methods in the following section that takes the fitness between the computed and measured data into account.

	\vspace{-0.4cm}
	\section{Learning with physics loss}
	\label{sec:physics loss}
	
	This section presents several approaches that impose additional physical constraints on the loss function when training the network weights, different from conventional ML models using only the difference between predicted and labeled images as a loss function.  
	The advantage of additional constraints in loss is demonstrated in Box 1.
	
	\begin{spacing}{1.3}
		\begin{tcolorbox}[float,
			toprule = 0mm,
			bottomrule = 0mm,
			leftrule = 0mm,
			rightrule = 0mm,
			arc = 0mm,
			fonttitle = \sffamily\bfseries\large,
			title = Box1: Incorporating forward modeling in loss: a mathematical example]	
			
			Optimizing DNN parameters with the constraint of a forward problem can alleviate the ambiguity caused by the non-uniqueness of the inverse problem. 
			
			\ \ \ \  We demonstrate it using a toy problem \cite{shahriari2021error}, where the forward process has analytical solutions: \vspace{-0.6cm}
			\begin{equation}\label{key}
				m:= F(p) = p^2. \vspace{-0.4cm}
			\end{equation}
			The inversion has two branches of solutions: $ p=+\sqrt{m} $ and $ p=-\sqrt{m} $, see the black lines in Fig.~\ref{fig:fig5}. To solve the inverse problem with a DNN, the training dataset is constructed such that for each sample ($ m $,$\sqrt{m}  $), there is another one ($ m $,$-\sqrt{m}  $) in it. The point that simultaneously minimizes the distance between the two solutions is zero. When training is supervised by labels  $ p $ ($ \pm \sqrt{m} $), the predictions are zeros, see Fig.~\ref{fig:fig5} (left), which are fake answers caused by the non-uniqueness. On the other hand, when training is supervised by labels $ p^2 $, the correct branch can be predicted by controlling the signs of solutions, see Fig.~\ref{fig:fig5} (right). In EM imaging, the forward problems are described by Maxwell's equations.
			\\ 
			\centerline{
				\begin{minipage}[t]{\linewidth}        
					\includegraphics[height=50mm]{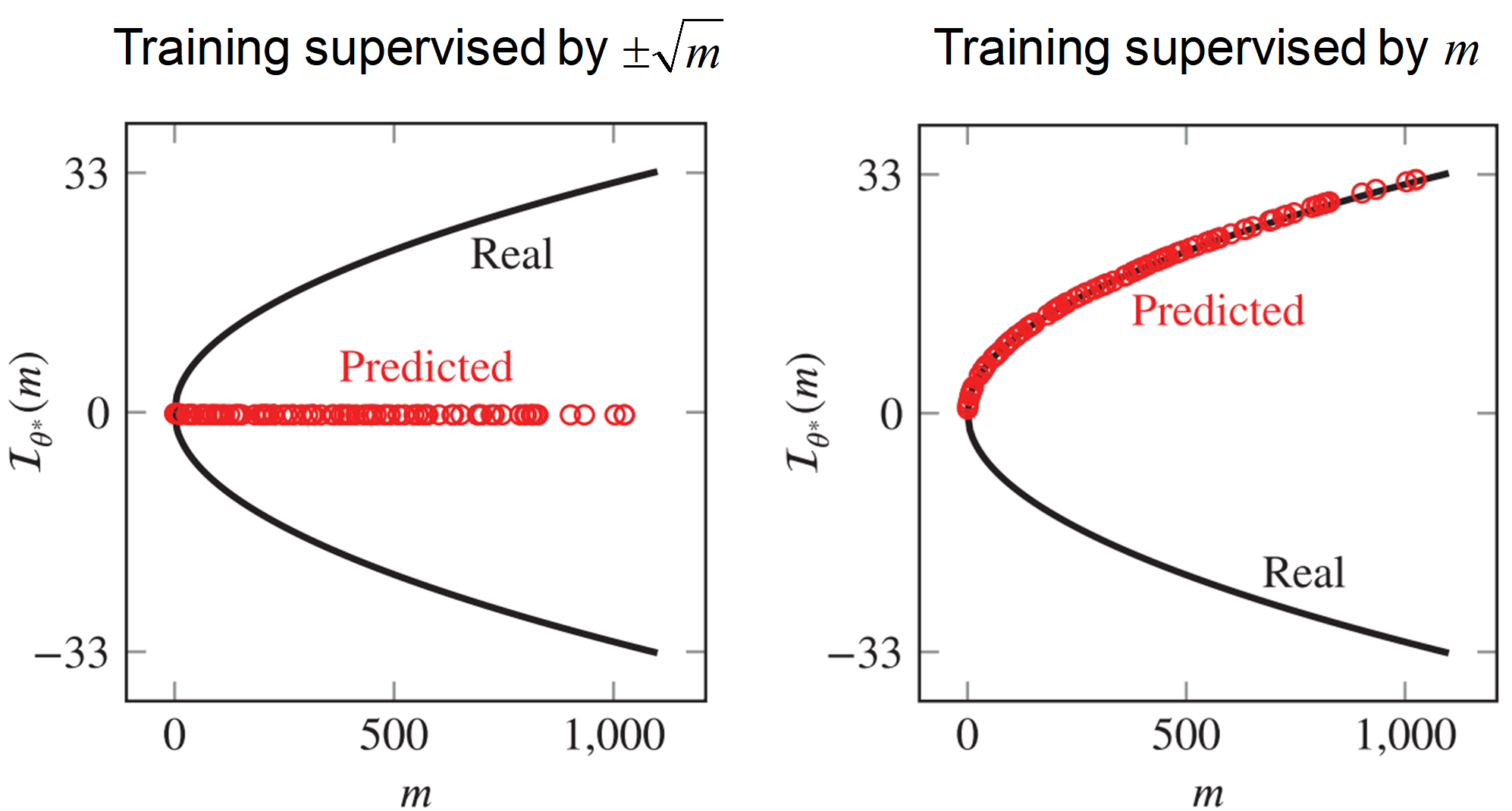}\vspace{-0.4cm}
					\centering \captionof{figure}{Incorporating forward modeling into training to reduce the non-uniqueness of the inverse problem.}\label{fig:fig5}
			\end{minipage}}
			
			
		\end{tcolorbox}		
	\end{spacing}
	
	In the following, we consider different types of physics losses: rigorous measurement loss, learned measurement loss, and PDE-constrained loss. 
	
	\vspace{-0.4cm}
	\subsection{Training with a rigorous measurement loss}
	Consider the inverse problem solved by a DNN with the measured data $ \mathbf d $ as input and the permittivity $ \boldsymbol{\epsilon} $ as output. When the EM data at the receivers can be numerically computed as in many applications,  one can  embed the data fitness, which involves physical rules, in the training loss function  \cite{jin2020physics}. 
	
	Let  $ \boldsymbol{\epsilon}_T $ and $ \mathbf d_T$ denote the labeled permittivity and EM data for training, respectively. Purely data-driven imaging uses permittivity loss $ L_{\epsilon} = \|\boldsymbol{\epsilon} - \boldsymbol{\epsilon}_T \|^2  $ for training. The physics embedded one further incorporates the measurement (data) loss $ L_\text{d} $, given by
	\vspace{-0.4cm}
	\begin{equation}\label{eq14}
		L = \alpha L_{\epsilon} + \beta L_\text{d} = \alpha \|\boldsymbol{\epsilon} - \boldsymbol{\epsilon}_T \|^2 + \beta \| F(\boldsymbol{\epsilon}) - \mathbf d_T \|^2, \vspace{-0.4cm}
	\end{equation}
	where  $ \alpha $ and $ \beta $ are weighting coefficients. If $ \alpha=0 $, the DNN training can be regarded as unsupervised learning. {\color{black}In the geosteering EM data inversion\cite{jin2020physics}, this scheme achieves two orders of magnitude lower data misfit compared with training with permittivity loss only ($ \beta=0 $).  } 
	
	Training such a DNN requires backpropagating the gradients of the data misfit, where the Fréchet derivative  $ \partial F/ \partial \boldsymbol{\epsilon} $ needs to be computed outside the DL framework. Methods for estimating the derivative have been addressed in traditional deterministic inversion, such as the finite difference method or adjoint state method  \cite{abubakar20082}. 
	
	\vspace{-0.4cm}
	\subsection{Training with a learned measurement loss}
	Computing the Fréchet derivative  $ \partial F/ \partial \boldsymbol{\epsilon} $ is time-consuming. 
	An idea to accelerate it is to surrogate the numerical forward solver $F(\cdot)$ with a DNN $\Theta_F(\cdot)$ \cite{shahriari2021error}. 
	The training contains two stages: 1) training the forward solver $ \Theta_F^* $ and 2) training the inverse operator $ \Theta_I^* $, given by \vspace{-0.3cm}
	\begin{equation}\label{key}
		\Theta_F^* = \arg \min _{\Theta_F}\left\|\Theta_F(\boldsymbol{\epsilon}_T)-\mathbf{d}_T\right\|^2, \quad	\Theta_I^* = \arg \min _{\Theta_I}\left\|\Theta_F^*  (\Theta_I\left(\mathbf d_T\right))-\mathbf d_T\right\|^2. \vspace{-0.2cm}
	\end{equation}
	Both stages take measurement misfit as the loss function, which involves physical rules. 
	
\begin{figure}[!]
		\centering
		\includegraphics[width=100mm]{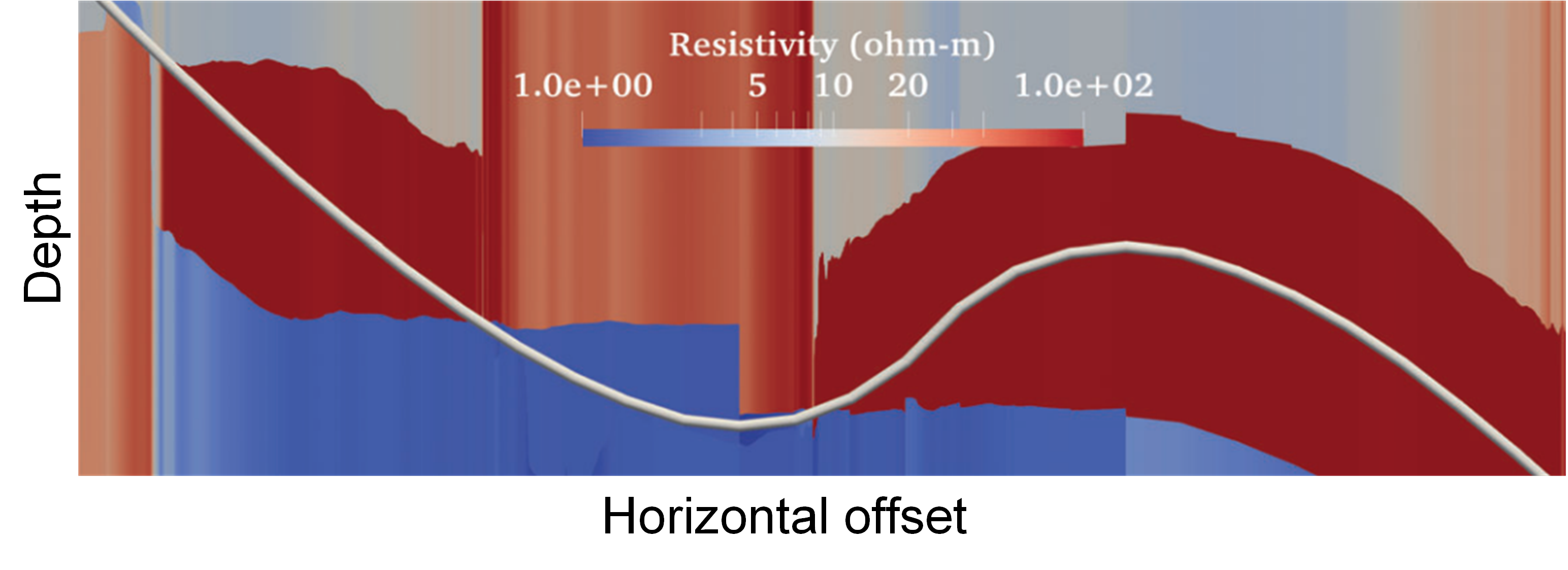}	
		\caption{Underground resistivity predicted from logging-while-drilling EM data \cite{shahriari2021error}. The gray curve represents the drilling trajectory, along which EM fields are transmitted and collected by a logging instrument. The resistivity around the trajectory is recovered by a DNN trained with a learned measurement loss.     }
		\label{figdwl}
	\end{figure}

	This scheme successfully solves the logging-while-drilling inverse problem for borehole imaging \cite{shahriari2021error}, see Fig.~\ref{figdwl}, where the purely data-driven approach did not achieve satisfactory reconstructions due to the severe nonlinearity and ill-posedness in logging-while-drilling inversion \cite{shahriari2020deep}. 
	
	\vspace{-0.4cm}
	\subsection{Training with a PDE-constrained loss}
	The PDE-constrained loss inserts partial differential equations (PDEs) into the loss function. The representative work is the physics-informed neural network (PINN) \cite{raissi2019physics,karniadakis2021physics} that is designed for both forward and inverse problems. It is a mesh-free method and can seamlessly fuse knowledge from observations and physics. We present an example of PINN for the inverse problem in Box 2. 
	
	\begin{spacing}{1.3}
		\begin{tcolorbox}[float,
			toprule = 0mm,
			bottomrule = 0mm,
			leftrule = 0mm,
			rightrule = 0mm,
			arc = 0mm,
			fonttitle = \sffamily\bfseries\large,
			title = Box2: Physics-informed neural network (PINN) for the inverse problem]	
			
			Consider the one dimensional time-domain EM wave equation \vspace{-0.1cm} 
			\begin{equation}\label{eq6}
				\frac{\partial^{2} E(x,t)}{\partial x^{2}}-\mu \epsilon(x) \frac{\partial^{2} E(x,t)}{\partial t^{2}}=0, \vspace{-0.1cm}
			\end{equation}
			where $ E $ is the electric field, $ \epsilon $ is permittivity, $ \mu $ is permeability, $ t $ and $ x $ is the time and spatial coordinate, respectively. Together with some boundary conditions, the equation can be analytically or numerically solved to yield $E$ (forward problem) or yield $\epsilon$  (inverse problem) given $t$ and $x$.  
			
			\ \ \ \ Take the inverse problem with one-source-multiple-receivers as an example. A PINN specifies two separate DNNs, namely $ \Theta_F $ and $ \Theta_I $. The input of $ \Theta_F $ is $ x $ and $ t $ and its output is the electric field $ \tilde{E} $, denoted by $\tilde{E}=\Theta_F(x,t) $. Similarly, the input of $ \Theta_I $ is $ x $ and its output is permittivity $ \tilde{\epsilon} $, denoted by $ \tilde{\epsilon}=\Theta_I(x) $. The two separate DNNs are simultaneously trained with a shared loss function $L $, which includes a supervised measurement loss of $ E $ regarding initial and 
			boundary conditions  \vspace{-0.2cm}
			\begin{equation}\label{key}
				{L}_{\text {data }}=\frac{1}{N_{\text {data }}} \sum_{i=1}^{N_{\text {data }}}\left(\tilde{E}\left(x_{i}, t_{i}\right)-{E}_T\left(x_{i}, t_{i}\right)\right)^{2}\vspace{-0.2cm}
			\end{equation}\vspace{-0.1cm}
			and an unsupervised loss of PDE constructed according to (\ref{eq6}) \vspace{-0.2cm}
			\begin{equation}\label{key}
				{L}_{\mathrm{PDE}}=\frac{1}{N_{\mathrm{PDE}}} \sum_{j=1}^{N_{\mathrm{PDE}}}\left(\frac{\partial^{2} \tilde{E}\left(x_{j}, t_{j}\right)}{\partial x^{2}}-\mu \tilde\epsilon(x_j) \frac{\partial^{2} \tilde{E}\left(x_{j}, t_{j}\right)}{\partial t^{2}}\right)^{2}, \vspace{-0.1cm}
			\end{equation}
			given by
			$L = \alpha_\text{data} L_\text{data} + \alpha_\text{PDE} L_\text{PDE}$. 				
			Here  $ \left(x_{i}, t_{i}\right) $ and $  \left(x_{j}, t_{j}\right) $ is sampled at the initial/boundary position and in the \ac{doi}, respectively. In addition, $ E_T $ is the labeled measurement, $ N_\text{data}$ is the number of labeled samples, $   N_\text{PDE}$ is the number of unlabeled samples in the \ac{doi}, and $  \alpha_{\cdot} $ are weights. 
			The partial differentiations are achieved by the automatic differentiation in the DL framework. After training, one can use $ \Theta_I $ to predict permittivity at arbitrary location $ x $. Therefore, the PINN is mesh-free.
		\end{tcolorbox}		
	\end{spacing}

	PINN is applied to electrical impedance tomography in \cite{bar2021strong}. Electrical impedance tomography measures the voltages on a body surface after electric currents are injected. The imaging recovers conductivity distribution inside the body. It usually contains multiple transmitting and receiving sensors. In Box 2, we show that a PINN contains one forward network $ \Theta_F $ and one inverse network $ \Theta_I $ for one transmitting source. When J sources illuminate the domain, the PINN will contain J+1 networks, corresponding to J forward networks that output voltages generated by different sources and one inverse network that outputs conductivity. Furthermore, the loss function should be modified to $ L = \sum_i^J {L_i} $, where $ L_i $ represents the loss function for the $ i $-th source. Simultaneously training all networks can satisfy both PDEs and boundary conditions (measurements).
	
	The smoothness of conductivity and known conductivity on the boundary are represented as regularizations in the loss function of PINN to stabilize the inverse process \cite{bar2021strong}.  In numerical simulations, the authors set J=8, $ N_\text{data}$=10,000 and $ N_\text{PDE}$=8000, and achieve better results than two conventional methods. However, one should notice that we seldom have so many measurements in reality, so its performance on experimental data imaging needs to be further investigated.
	
	\vspace{-0.5cm}
	\subsection{Discussions}
	When inverting limited-aperture EM data, insufficient measurements may lead to the instability of training a PINN. In this case, it would be better to use the first two approaches that explicitly define the measurement loss at the receivers. PINN outperforms the two approaches when simulating the EM response is prohibitive, for example, due to the high computational cost or complex EM environment. Finally, when recovering diverse targets, the former two approaches can make predictions without retraining the neural network, while PINN needs to be trained {\color{black} for each target}. 
	
	\vspace{-0.2cm}
	\section{Learning with physics models}
	\label{sec:unroll}
	\vspace{-0.1cm}
	Following the use of unrolling in other domains \cite{8434321,sahel2022deep,monga2021algorithm}, unrolling has also been used in EM imaging models, yielding physics embedded neural networks. We group this type into three subtypes: unrolling the measurement-to-image (inverse) mapping, unrolling the image-to-measurement (forward) mapping, and simultaneously unrolling both mappings.
	\vspace{-0.4cm}
	\subsection{Unrolling measurement-to-image mapping}
	\subsubsection{Linear problem}
	We demonstrate the unrolling of linear inverse problems through radar imaging. \Revise{Here, the electric parameters of interest are intensities of scatterers in the \ac{doi}, denoted by $\boldsymbol \epsilon$ with slight abuse of notation. Then, linear approximation is usually applied in the forward model for high computational efficiency, yielding $F({\boldsymbol{\epsilon}})=\bm \Phi {\boldsymbol{\epsilon}}$, where $\boldsymbol \Phi$ is a matrix determined by the radar waveform and the geometry of the \ac{doi}.} Conventional radar imaging can be formulated as a compressed sensing problem:  $\min_{\boldsymbol{\epsilon}} \|\mathbf d_\text{obs} - \bm \Phi {\boldsymbol{\epsilon}}\|^2 + \lambda \left\| \boldsymbol{\epsilon} \right\|_1$. There are a myriad of methods proposed to solve such problems. A well-known  technique is the \ac{ista} that iteratively performs proximal gradient descent \cite{Beck2009A}. Specifically, the solution is updated by:
	\vspace{-0.2cm}
	\begin{eqnarray}\label{eq:ISTA}
		\boldsymbol{\epsilon}_{k}  = {\mathcal{S}_{\frac{\lambda }{L}}}\left( {\frac{1}{L}{\bm \Phi ^H} {\mathbf d}_\text{obs} + \left( {\bm I - \frac{1}{L}{\bm \Phi ^H}\bm \Phi } \right){\boldsymbol{\epsilon}_{k-1}}} \right), \vspace{-0.4cm}
	\end{eqnarray}
	where $L = {\lambda _{\max }}( {{\bm \Phi ^H}\bm \Phi } )$ is the Lipschitz constant, and $\lambda _{\max }(\cdot)$ represents the maximum eigenvalue of a Hermitian matrix, {\color{black}and $\cdot^H$ denotes the conjugate transpose}. The element-wise soft-threshold operator {\color{black}$\mathcal{S}_{\theta}$ assigns those elements below the threshold $\theta$ to zeros}, defined as 
	$\left[ {\mathcal{S}_{\theta} }\left( \bm{u} \right) \right]_i= \text{sign} \left( {{[\bm{u}]_i}} \right)\left( {\left| {{[\bm{u}]_i}} \right| - \theta} \right)_ +
	$, where sign$ (\cdot)$ returns the sign of a scalar, $ (\cdot)_+ $ means $ \max(\cdot,0)$, 
	and $ \theta $ is the threshold.  \ac{ista}  shows high accuracy but requires thousands of iterations for convergence. 
	
	To accelerate the solution process, \ac{lista} with only several neural network layers is proposed in \cite{Gregor2010Learning}, where each layer  unfolds an \ac{ista} iterations. Particularly, \ac{lista} treats ${\lambda /L}$, ${\frac{1}{L}{\bm \Phi ^H}}$  and ${\left( {\bm I - \frac{1}{L}{\bm \Phi ^H}\bm \Phi } \right)}$ in \eqref{eq:ISTA} as  variables to learn from training data with a back-projection algorithm, disregarding their physics structures. Numerical results in \cite{Gregor2010Learning} show that  \ac{lista} can achieve virtually the same accuracy as \ac{ista} using nearly two-order fewer iterations and does not require knowledge of $ \bm \Phi $. Nevertheless, a challenge in \ac{lista} is that there are many variables to learn, requiring carefully tuning of hyper-parameters to avoid over-fitting and gradient vanishing.

	Embedding physics models into the neural networks reduces the number of variables while maintaining fast convergence rate \cite{Fu2021Toep,fu2022block}. For example, the mutual inhibition matrix $\bm I - \frac{1}{L}{\bm \Phi ^H}\bm \Phi $ has a Toeplitz or a doubly-block Toeplitz structure due to the nature of radar forward models. The degrees of freedom with such Toeplitz structure are reduced to $O(N)$ from $O(N^2)$, the counterpart without this structure. By incorporating such structure, the proposed method in \cite{Fu2021Toep} significantly reduces the dimension of neural networks, thereby reducing the amount of training data, memory requirements,  and computational cost, while maintaining comparable imaging quality as \ac{lista}. A similar approach is also adopted in \cite{fu2022block}, which explores the coupling structure between different blocks in the radar forward model $\bm \Phi$.

	\subsubsection{Nonlinear problem}
	The objective function of nonlinear EM imaging, $	L(\boldsymbol{\epsilon}) = \| \mathbf d_\text{obs} - F(\boldsymbol{\epsilon})\|^2$, where $ F(\boldsymbol{\epsilon}) $ is numerically solved from PDEs, is conventionally minimized through gradient descent methods. With the Gauss-Newton method, permittivity is updated according to
	\vspace{-0.4cm} 
	\begin{equation}\label{eq13}
		\boldsymbol{\epsilon}_{k+1} = \boldsymbol{\epsilon}_k+ (\mathbf S^H \mathbf S)^{-1}\mathbf S^H \big( \mathbf d_\text{obs} - F( \boldsymbol{\epsilon}_{k}) \big), \vspace{-0.4cm}
	\end{equation}
	where $ \mathbf S $ is the Fréchet derivative of $ F $ at $  \boldsymbol{\epsilon}_0 $. Notice that  $ \mathbf S $ only contains the local property of the objective function, and computing $ F( \boldsymbol{\epsilon}) $, $ \mathbf S $ and $ (\mathbf S^H \mathbf S)^{-1} $ is usually expensive.
	
	By unrolling, a set of descent directions $ \mathbf Ks $ can be learned instead of computing $ \mathbf S $ and $ (\mathbf S^H \mathbf S)^{-1}\mathbf S^H $ online. This is called the supervised descent method (SDM), which was first proposed for solving nonlinear least-squares problems in computer vision \cite{xiong2013supervised}. In online imaging, permittivity is updated by
	\vspace{-0.4cm}
	\begin{equation}\label{key}
		\boldsymbol{\epsilon}_{k+1} = \boldsymbol{\epsilon}_k + \mathbf K_k \big( \mathbf d_\text{obs} - F(\boldsymbol{\epsilon}_k) \big).\vspace{-0.4cm}
	\end{equation}
	In training, the EM response is taken as the input, while the corresponding ground truth of {\color{black}complex} permittivity is the label. More details on the training can be found in \cite{guo2020pixel}.
	
\begin{figure}[!]
		\centering
		\includegraphics[width=162mm]{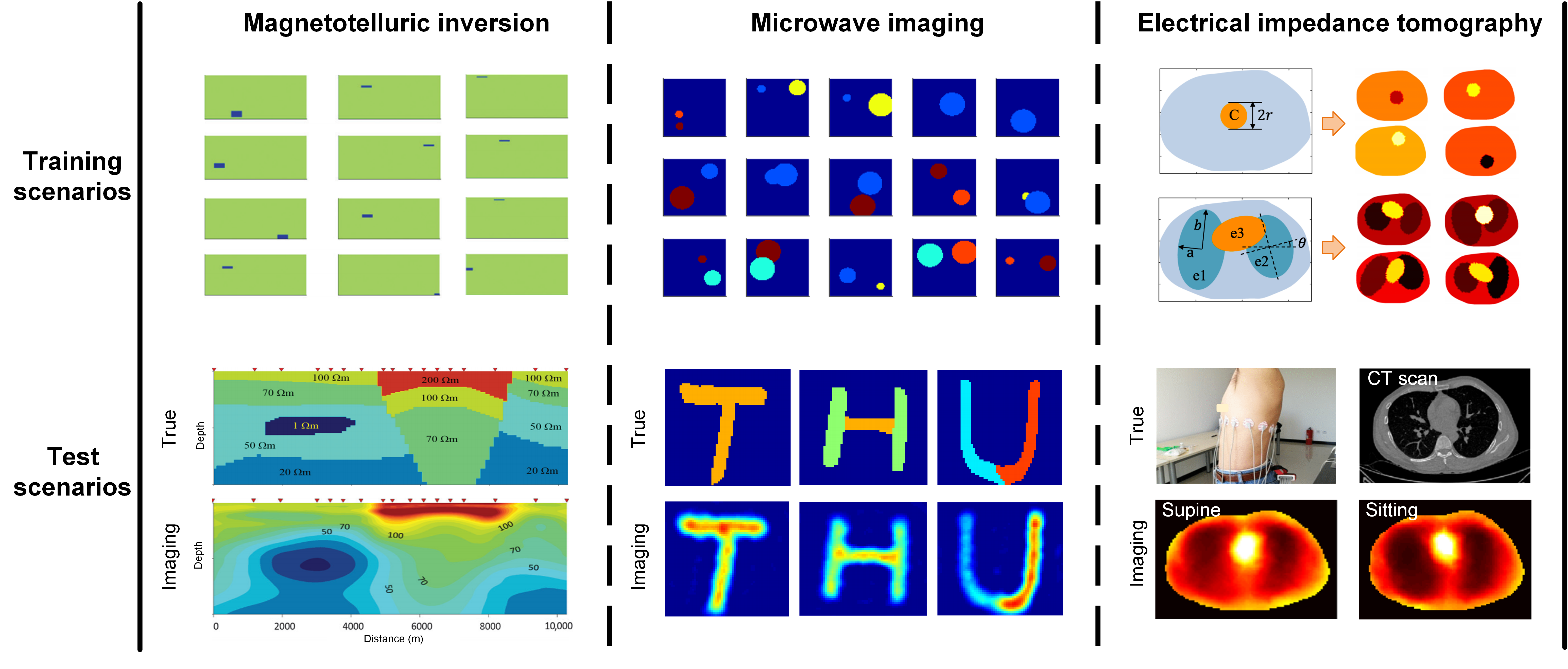}	
		\caption{Imaging with SDM for geophysics \cite{guo2020application}, microwave \cite{guo2020pixel} and biomedical data \cite{zhang2020supervised}. SDM is able to reconstruct complex inhomogeneous media while the training scenarios are simple. }
		\label{fig6}
	\end{figure}

	SDM shows high generalizability in EM imaging. 
	The imaging process is mainly governed by physical law, with a soft constraint imposed by the learned descent directions. Experiments show that the descent directions trained with simple scatterers can be applied to predict complex targets in inhomogeneous media for various applications, such as geophysical inversion \cite{guo2020application,hu2020supervised,lu20211,hao2021robust}, microwave imaging \cite{jia20213,guo2020pixel}, and biomedical imaging \cite{lin2020neural,zhang2020supervised}. Fig.~\ref{fig6} shows its applications and generalizability.
	
	SDM may be flexibly combined with techniques in conventional gradient-based inversion. For example, using regularizations on the objective function of prediction, images can be predicted with either smooth \cite{guo2020application} or sharp \cite{guo2019regularized} interfaces without retraining descent directions. Furthermore, SDM and conventional  methods can be flexibly switched to satisfy different requirements of speed and accuracy  \cite{guo2020application}. 
	
	A limitation of SDM is the lower speed of prediction compared with end-to-end DNNs, because the online forward modeling is in general computationally intensive. Efforts have been made to unroll the nonlinear forward modeling to accelerate imaging, which will be discussed in the following.
	\vspace{-0.4cm}
	\subsection{Unrolling image-to-measurement mapping}
	{\color{black} Accelerating the forward process also improves the efficiency of EM imaging. This part introduces two methods where the frequency-domain and time-domain forward modeling is accelerated by unrolling integral and differential operations, respectively. }
	\subsubsection{\color{black}Unrolling the integral operation}
	In this scheme, the differential equation (\ref{eq1}) is first converted into an integral form, and then the forward modeling $ F(\boldsymbol{\epsilon}) $ involving integral operations is unrolled as a physics embedded network $ \Theta_F $. After the networks are trained, they are combined with generic networks  $ \Theta_I $  that perform inverse mappings to achieve full-wave EM imaging; {\color{black}hence the cascaded neural networks  becomes  a physics embedded DNN (PE-Net) \cite{guo2021physics2}}.
	
\begin{figure}[!]
		\centering
		\includegraphics[width=145mm]{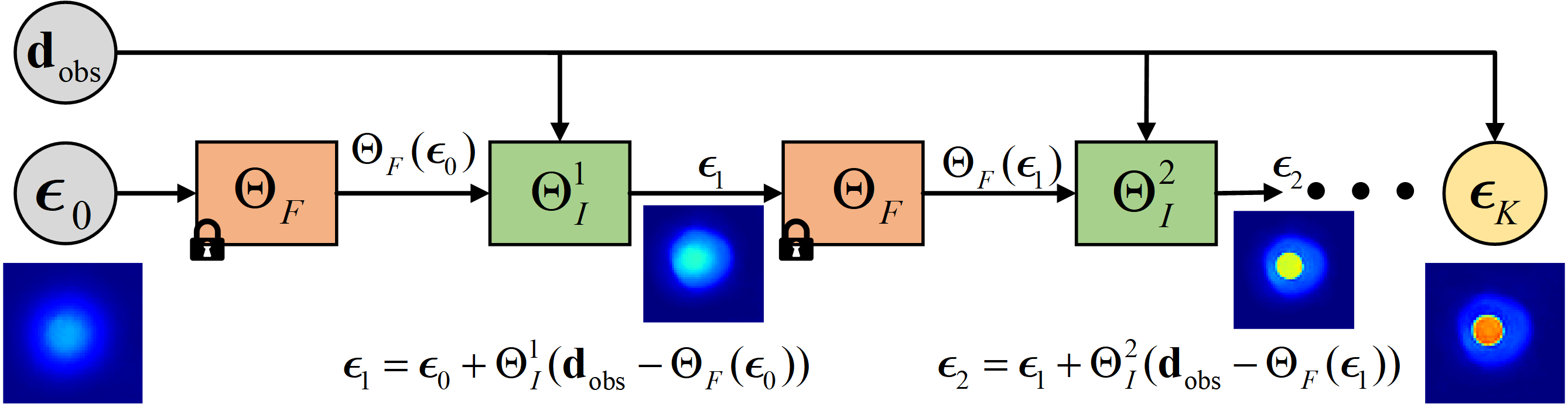}	
		\caption{Physics embedded DNN for microwave imaging \cite{guo2021physics2}, where the forward modeling is unrolled into a neural network. The parameters of the forward solver $\Theta_{F}$ are fixed when training the inverse networks $\Theta_{I}$s.  }
		\label{fig9}
	\end{figure}

	{\color{black}The schematic architecture of the PE-Net is shown in Fig.~\ref{fig9}, where $ {\boldsymbol{\epsilon}}_0 $ is the initial permittivity, $ \Theta_F $ represents the DNN-based forward modeling solver, and $\{ \Theta_I^1, \Theta_I^2,\dots \} $ represent {\color{black}neural networks} that predict the update of {\color{black}complex} permittivity. Let $ K $ be a predefined maximum number of iterations. The final output $ \boldsymbol{\epsilon}_K $ is then }
	\vspace{-0.4cm}
	\begin{equation}\label{key}
		\boldsymbol{\epsilon}_{K}=\Theta_{I}\left( {\boldsymbol{\epsilon}}_0,  {\mathbf{d}}_\text{{obs }}\right)= {\boldsymbol{\epsilon}}_{0}+\sum_{k=1}^{K} \Theta_{I}^{k}\left({\mathbf{d}}_\text{{obs }}-\Theta_{F}\left({\boldsymbol{\epsilon}}_{k-1}\right)\right),\vspace{-0.3cm}
	\end{equation} 
	where $ k $ is the index of iterations.
	
	The integral form of the wave equation is \vspace{-0.3cm} 
	\begin{equation}\label{vie}
		\mathbf E(\mathbf r) = \mathbf E^\text{inc}(\mathbf r)  + \omega^2\mu\int_{V} \vec{\vec{\mathbf G}}_0\left(\mathbf{r}, \mathbf{r}^{\prime}\right) \left[\epsilon(\mathbf r^\prime) - \epsilon_0\right] \mathbf{E}(\mathbf{r^\prime}) d\mathbf r^\prime, \vspace{-0.3cm} 
	\end{equation}
	where $ \mathbf E^\text{inc} $ is the incident field generated by the source, $\vec{\vec{\mathbf G}}_0$ is the Green's function describing wave propagation, $ \epsilon_0 $ is the permittivity of the background, and $ V $ is the \ac{doi}.
	Here, $ \Theta_F $, which essentially solves {(\ref{vie})}, is established by unfolding the conjugate gradient method. {\color{black}Conventionally, solving (\ref{vie}) is simplified as calculating $ \mathbf x $ (representing the unknown $ \mathbf E $) from $ \mathbf A (\boldsymbol{\epsilon}) \mathbf x = \mathbf b $, where $ \mathbf A(\boldsymbol{\epsilon}) $ is a matrix related to wave physics and target permittivity, and $ \mathbf b $ is a constant vector. Generally,  $ \mathbf A $ is a full matrix with millions of elements, making solving with iterative matrix equation solvers very time-consuming. In \cite{guo2021physics1}, inspired by the conjugate gradient method, the matrix equation is solved by alternately predicting the conjugate direction $ {\mathbf p} $ and the solution update $ \Delta{\mathbf x} $ iteratively. Details can be found in Box 3. 
	 Experiments show that $ \Theta_F $ {\color{black}needs much less iterations} than the conjugate gradient method.} 
	
	\begin{spacing}{1.3}
		\begin{tcolorbox}[float,
			toprule = 0mm,
			bottomrule = 0mm,
			leftrule = 0mm,
			rightrule = 0mm,
			arc = 0mm,
			fonttitle = \sffamily\bfseries\large,
			title = Box3: Conjugate gradient method and update-learning method]
			\setlength{\columnsep}{30pt}
			{\color{black}To solve the matrix equation $ \mathbf A \mathbf x=\mathbf b $ whose  $ \mathbf A $ is positive-definite, the conjugate gradient method computes the conjugate direction $ \mathbf p $ and updates the solution in an iterative manner. The solution process may take hundreds to thousands iterations. 
			
			\ \ \ \  In EM modeling, incident waves at a certain frequency behave similarly, which leads to a limited diversity of the right-hand term ${\mathbf b}$. In addition, the background Green's function does not change with different targets. Thus, characteristics of matrix equations derived from (\ref{vie}) are similar given a  certain scenario. This enables the machine to learn how to update ${\mathbf p}$ and ${\mathbf x}$ from training data, rather than computing them online as in the conjugate gradient method, to obtain a faster convergence speed.
			
			\ \ \ \ In the update-learning method\cite{guo2021physics2},  iterations in conjugate gradient method are unrolled by $ N $ neural network blocks, which contain cascaded neural network  $ \Theta_p $ and  $ \Theta_{dx} $ that predict the quasi conjugate direction and the solution update, respectively. Detailed algorithms of the two approaches are shown below.}
			\vspace{-0.4cm}
			\begin{multicols}{1}
				\begin{center}
					\begin{tabular}{p{7.0cm}c}
						\toprule
						{Conjugate gradient method}  \\        
						\midrule
						1: $\;$Input  $ {\mathbf x}_0$\\
						2: $\;$$ {\mathbf r}_0 = {\mathbf b} - {{\mathbf A}}{\mathbf x}_0$, ${\mathbf p}_1 =  {\mathbf r}_0$\\
						3: $\;$$\alpha_1=({\mathbf r}_0^T {\mathbf r}_0) / {\mathbf p}_1^T ({{\mathbf A}} {\mathbf p}_1)$\\
						4: $\;$${\mathbf x}_1 = {\mathbf x}_0 + \alpha_1 {\mathbf p}_1$\\
						5: $\;$\textbf{for} $\textit{k}$ = 1,2,..., \textbf{until} $|| {\mathbf r}_k|| \leqq \varepsilon$\\
						6: $\;$$\quad\quad {\mathbf r}_k= {\mathbf r}_{k-1} - \alpha_k ({{\mathbf A}}{\mathbf p}_k)$   \\
						7: $\;\quad\quad \beta_{k+1} = ({\mathbf r}_k^T {\mathbf r}_k)/({\mathbf r}_{k-1}^T {\mathbf r}_{k-1})$ \\
						8: $\;\quad\quad{\mathbf p}_{k+1} = {{\mathbf r}_k} + \beta_{k+1} {\mathbf p}_{k}$ \\
						9: $\;\quad\quad \alpha_{k+1}=({\mathbf r}_{k}^T {\mathbf r}_{k}) / {\mathbf p}_{k+1}^T ({{\mathbf A}} {\mathbf p}_{k+1})$\\
						10:$\;\quad\quad {\mathbf x}_{k+1} = {\mathbf x}_{k} + \alpha_{k+1} {\mathbf p}_{k+1}$ \\
						\bottomrule\\
					\end{tabular}
				\end{center}
				\begin{center}
					\begin{tabular}{p{7.0cm}c}
						\toprule
						{Update-learning method}  \\        
						\midrule
						1: $\;$Input  $ {\mathbf x}_0$\\
						2: $\;$$ {\mathbf r}_0 = {\mathbf b} - {{\mathbf A}}{\mathbf x}_0$, ${\mathbf p}_1 =  {\mathbf r}_0$\\
						3: $\; {\mathbf x}_1$ = $ {\mathbf x}_0$\\
						4: $\;$\textbf{for} $\textit{k}$ = 1,2,...,$N$,\\
						5: $\;$$\quad\quad {\mathbf r}_k= {\mathbf b} -  {{\mathbf A}}{\mathbf x}_k$   \\
						6: $\;\quad\quad{\mathbf p}_{k+1} = \Theta_p^k({\mathbf p}_{k}, {\mathbf r}_{k}, {\mathbf r}_{k-1})$\\
						7: 	$\;\quad\quad{\mathbf x}_{k+1} = {\mathbf x}_{k} + \Theta_{dx}^k({\mathbf p}_{k+1},   {{\mathbf A}} {\mathbf p}_{k+1}, {\mathbf r}_{k} )$ \\
						\bottomrule
					\end{tabular}
				\end{center}
			\end{multicols}
		\end{tcolorbox}
	\end{spacing}
	Aside from fast convergence, $ \Theta_F $ possesses high generalizability thanks to the incorporation of wave physics. Note that the $ \mathbf{A} $ matrix is constructed inside the network by integral operations, which involves Green's function that describes the interactions of electric fields in the entire domain. The explicit field integration brings global information of wave propagation to the receptive field of convolutional layers. {\color{black}After training with 32000 samples, the network can predict the field of targets statistically different from the training ones in real time, which builds the foundation for the following imaging problem.}
	
	The weights of the $ \Theta_{F} $ are fixed during training $ \Theta_I $s. The $ \Theta_I $s are achieved by generic networks, which performs the nonlinear mapping from the measurement domain to the permittivity domain. In training, EM fields generated by synthetic targets are taken as the input, while the target images are taken as labels. {\color{black}In prediction, the network achieves super-resolution reconstruction of targets that are quite different from the training ones. At the same time, the simulated field of the predicted target is in good agreement with the observed field.} 

	\subsubsection{Unrolling the differential operation}
	{\color{black}
		Attempts have also been made to unroll the time-domain wave equation with recurrent neural networks (RNNs) \cite{hu2021theory,guo2021electromagnetic}, where differential operations are represented by network layers.  In the 2D case, Maxwell's equations can be written as
		\begin{equation}\label{eq15}
			\left\{\begin{array}{c}
				\epsilon_R \frac{\partial E_{z}}{\partial t}=\frac{\partial H_{y}}{\partial x}-\frac{\partial H_{x}}{\partial y}-\sigma E_{z}, \\
				\mu \frac{\partial H_{x}}{\partial t}=-\frac{\partial E_{z}}{\partial y}, \\
				\mu \frac{\partial H_{y}}{\partial t}=\frac{\partial E_{z}}{\partial x},
			\end{array}\right. \vspace{-0.2cm}
		\end{equation}
		where $ E $ and $ H $ is the electric and magnetic field that are coupled with each other, the subscripts represent spatial components of the vector field, $ \epsilon_R $ is {\color{black}permittivity}, $ \sigma $ is conductivity, $ x $, $ y $, $ z $ and $ t $ is the spatial and time coordinates, respectively.
		
		After discretization, for instance, the $ \mu \frac{\partial H_{x}}{\partial t}=-\frac{\partial E_{z}}{\partial y} $ term becomes
		\begin{equation}
			\mu \frac{H_{x}^{n+1 / 2}\left(i, j+\frac{1}{2}\right) - H_{x}^{n-1 / 2}\left(i, j+\frac{1}{2}\right)}{\Delta t} = - \frac{E_{z}^{n}(i, j+1)-E_{z}^{n}(i, j)}{\Delta y}
		\end{equation}
		where $ (i,j+\frac{1}{2}) $, $ (i,j) $ and $ (i,j+1) $ represent the discrete spatial coordinate, $ n-1/2 $, $ n $ and $ n+1/2 $ represents the discrete time,  $ \Delta t $ and $ \Delta y $ are time and space intervals. The spatial differentiation in the right side can be represented by convolutional kernels. In the time domain, the $ H $-field at time $ n+1/2 $ is computed by $ H_{x}^{n+1 / 2} =  H_{x}^{n-1 / 2} + \Theta_F(E_{z}^{n})$, where $ \Theta_F $ represents basic operations on $ E_{z}^{n} $ realized by neural networks. Therefore, fields at the next time step can be updated from fields at current time steps. The updating process can be described by a recurrent neural network (RNN).
		
		\begin{figure}[!]
			\centering
			\includegraphics[width=110mm]{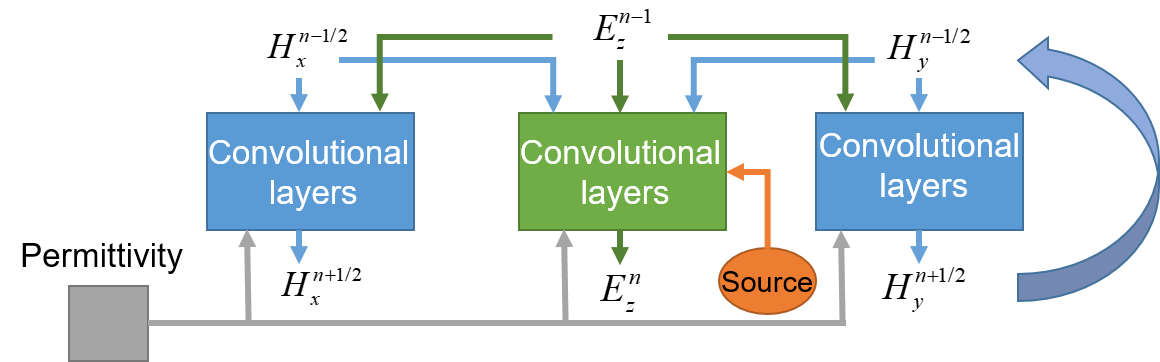}	
			\caption{{\color{black}The cell architecture of RNN for simulating wave propagation \cite{hu2021theory}. At each time step, the RNN outputs the $ E $-field $E_z$ and $ H $-fields $H_x$, $H_y$ in the entire \ac{doi}, which are computed from their values in the previous time step, according to Maxwell's equations. The partial derivatives are approximated with finite differences. Taking the permittivity as a trainable layer, training this network and updating its weights is equivalent to gradient-based EM imaging. }  }
			\label{fig8}
		\end{figure}
		
		Considering the couplings between $ E $- and $ H $-fields, the architecture unrolled from (\ref{eq15}) is shown in Fig.~\ref{fig8}. In the forward problem, after the target material is specified, EM data are generated by running the RNN without training. In the inverse problem, the material is represented by trainable parameters that are optimized by minimizing the misfit between simulated and labeled data. Training such a network and updating its weights is equivalent to gradient-based EM imaging. {\color{black}The use of automatic differentiation greatly improves the accuracy of gradient computation and achieves three orders of magnitude acceleration compared with the conventional finite difference method \cite{hu2021theory}.}
	}
	\vspace{-0.4cm}
	\subsection{Simultaneously unrolling both mappings}
\begin{figure}[!]
		\centering
		\includegraphics[width=120mm]{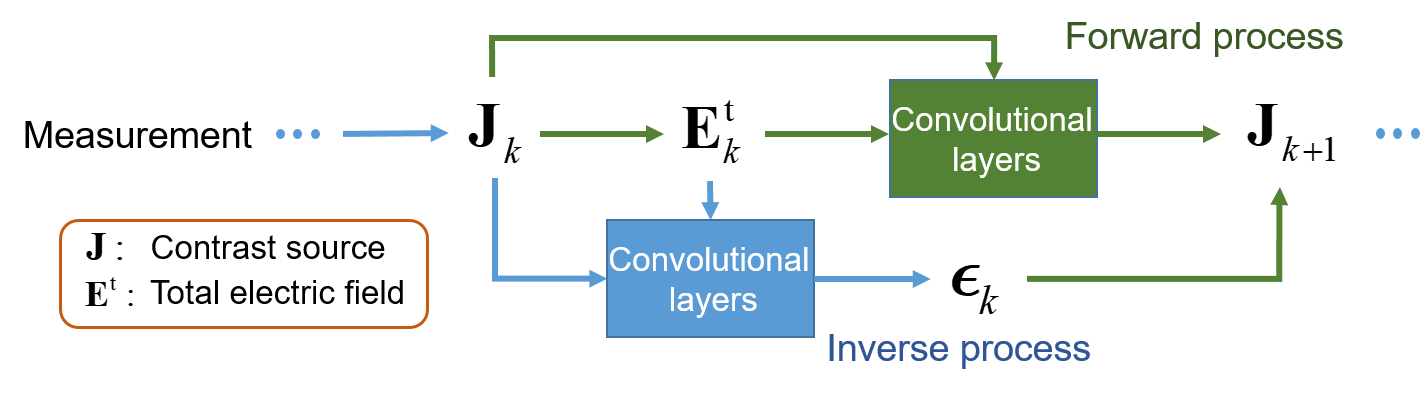}	
		\caption{{\color{black}Simultaneously unrolling forward and inverse processes into neural networks \cite{9539099}.  The forward process (in green) computes the contrast source $ \mathbf J $ and total field $ \mathbf E^\text{t} $ given permittivity, while the inverse process (in blue) infers permittivity from measurements,  $ \mathbf J $ and  $ \mathbf E^\text{t} $.
		}}
		\label{fig10}
	\end{figure}

	{\color{black}Many methods, such as the Born iterative method \cite{chew1990waves} or contrast source inversion \cite{van1997contrast}, solve the inverse scattering problem from a physics view: the target is progressively refined by simulating the physics more accurately. To be specific, they first reconstruct permittivity from a linear process by approximating the electric field $ \mathbf E $ in the integration of (\ref{vie}) to the incident field $ \mathbf E^\text{inc} $. Intermediate parameters, e.g., total field and contrast source, can be estimated with this permittivity. Then, a more accurate permittivity model is computed from the intermediate parameters and measurements, which are used to better approximate the intermediate parameters in the next iteration. When the intermediate parameters lead to scattered fields that fit measurements, the iteration stops and outputs the final estimated permittivity.
		
		The above process can be unfolded into neural networks \cite{9539099,shan2021neural}.  Here, for instance, the cell architecture of PM-Net \cite{9539099} is presented in Fig.~\ref{fig10}. Each cell contains approximated forward and inverse processes. In the forward process, the total field  $ \mathbf E $ is computed from the contrast source $ \mathbf J $, and a better contrast source is predicted from the total field and permittivity. In the inverse process, permittivity is predicted from the contrast source and total field. The network is trained with 300 handwritten letter profiles. In prediction, it achieves real-time imaging within 1 second while the alternating direction method of multipliers (ADMM) needs 300 seconds. The resolution is also largely improved \cite{9539099}.
	}
	
	\vspace{-0.4cm}
	{\color{black}
		\subsection{Comparisons}
		Based on \cite{9539099} and our previous work \cite{guo2021physics1}, it is possible to compare different strategies of deep unrolling when they solve the same inverse scattering problem\footnote{\color{black}We note that some simulation setups, e.g., electric scale, number of measurements, random noises, and training set, are not exactly the same. However, these differences are at an acceptable level and do not affect the conclusions of the comparison. }. Test data are simulated from the ``Austria" model that is often used as benchmark model in inverse scattering. It is challenging due to strong scatterings inside and among the rings. The reconstructed images with different approaches are presented in Fig.~\ref{fig11}, where ADMM is taken as the benchmark. The result in CS-Net is recovered by gradient-based optimization whose initial guess is provided by a DNN \cite{sanghvi2019embedding}.  BPS is in the scope of \textit{learning after physics processing}, where the neural network performs image enhancement after traditional qualitative imaging \cite{wei2018deep}. SDM unrolls the inverse mapping \cite{guo2020pixel}, PE-Net unrolls the forward mapping \cite{guo2021physics1} and PM-Net unrolls both mappings \cite{9539099}. In general, learning-governed methods can achieve higher resolution than physics-governed methods, thanks to incorporating prior knowledge through offline training. BPS has lower accuracy than the other three \textit{learning with physics model} approaches. PM-Net obtains the best accuracy by elaborately tailoring the neural network according to EM theory.
		
		\begin{figure}[!]
			\centering
			\includegraphics[width=60mm]{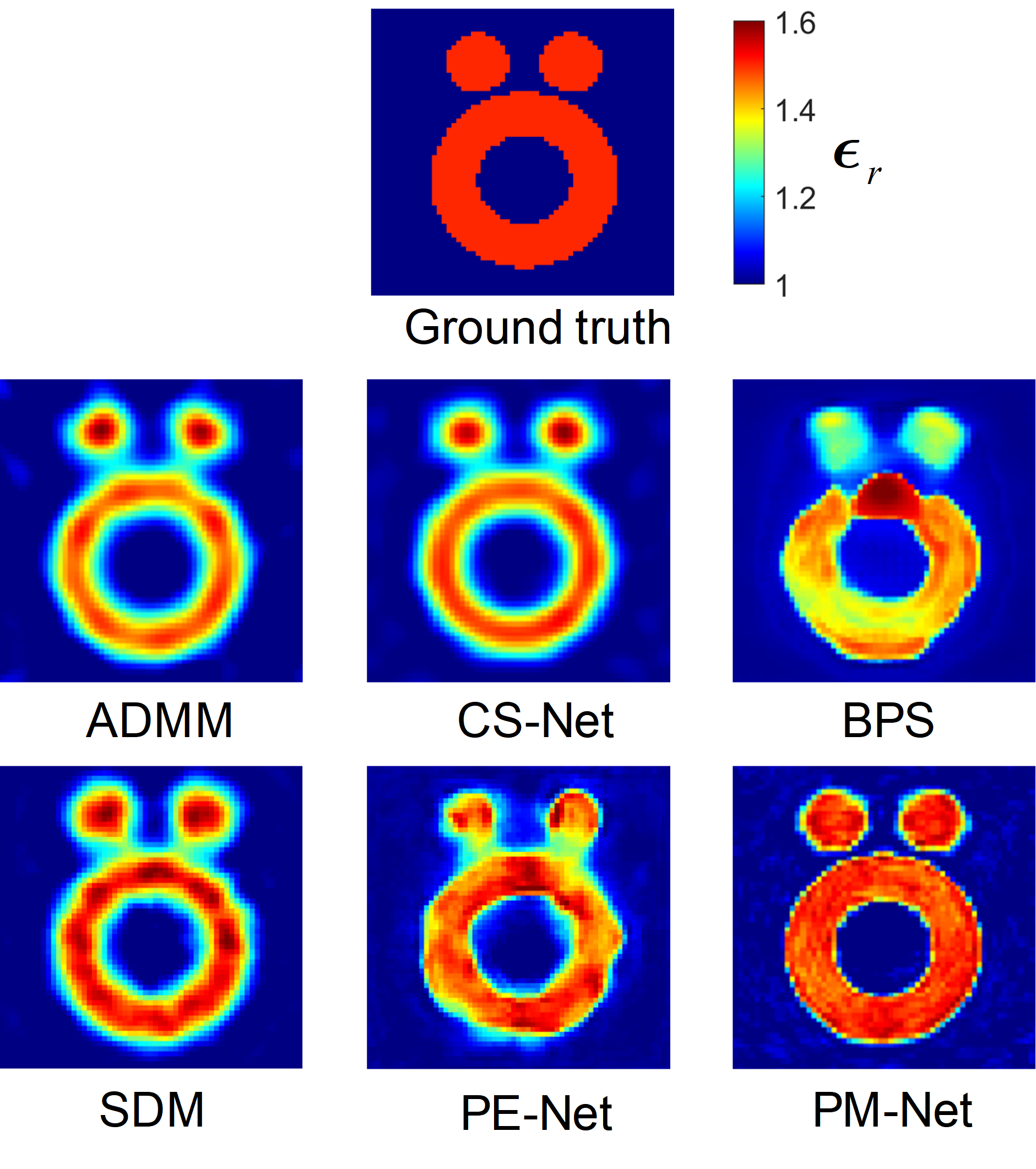}	
			\caption{{\color{black}Comparisons of the reconstructed ``Austria'' model with different methods  \cite{9539099,guo2020pixel,guo2021physics1}. The true relative permittivity ($\epsilon_r= {\epsilon}/{\epsilon_{0}} $) is 1.5. Imaging results are plotted in the same color scale. ADMM is taken as the benchmark. CS-Net \cite{sanghvi2019embedding} uses a neural network to provide initial guesses for gradient-based optimization. BPS   \cite{wei2018deep} belongs to \textit{learning after physics processing}. SDM, PE-Net and PM-Net belong to \textit{learning with physics models}. Specifically, SDM \cite{guo2020pixel} learns the inverse mapping, PE-Net \cite{guo2021physics1} unfolds the forward modeling, and PM-Net unfolds both forward and inverse mappings \cite{9539099}.	}}
			\label{fig11}
		\end{figure}
		
		Comparisons of imaging time and memory of network parameters are shown in Tab.~\ref{tab1}. Note that the SDM and PE-Net are trained for two-frequency and three-frequency imaging, respectively, while other methods are trained for single-frequency imaging. Therefore, the time/memory cost in SDM and PE-Net for single-frequency imaging can be less than the shown values. The imaging time of PE-Net and PM-Net is at the same level, much faster than other methods. CS-Net sequentially performs network inference and gradient-based optimization; hence it takes the most time. BPS improves  imaging speed by avoiding the EM modeling process. We note that the speed of BPS in the original paper \cite{wei2018deep} is faster than the presented one \cite{9539099}. SDM needs to rigorously solve Maxwell's equations in the prediction stage, hence it is slower than BPS, PE-Net and PM-Net. In addition, the memory of neural network parameters is also compared. Smaller networks imply that less training data are needed. SDM is the most memory-consuming, since it records several descent directions represented by full matrices. PM-Net requires the second-largest amount of memory since it unrolls a number of complete EM modeling processes into the neural network. In contrast, according to the inverse scattering theory, PM-Net finds a better way to unroll approximated forward and inverse problems, which leads to a much smaller network size.

		\begin{table}[!]
			\caption{\color{black}Imaging time and memory of network parameters in different methods  \cite{9539099}.  }
			\centering
			$\begin{array}{|c|c|c|c|c|}
				\hline \text { Methods } & \text { Imaging time (seconds)} & \text { Memory of parameters } \\
				\hline \text { ADMM } & 308.76  & -  \\
				\hline \text { CS-Net } & 531.24  & 144 \text{ MB}  \\	
				\hline \text { BPS } & 7.12  & 15 \text{ MB}  \\		
				\hline \text { SDM } & 19.87  & 2.3 \text{ GB}  \\	
				\hline \text { PE-Net } & 1.55  & 483 \text{ MB}   \\	
				\hline \text { PM-Net } & 0.99  & 0.59 \text{ MB}  \\					
				\hline 
			\end{array}$
			\label{tab1}
		\end{table}
	}
	\vspace{-0.2cm}
	\subsection{Discussion}
	Using trainable layers to unroll linear imaging problems is time- and memory-efficient but only suitable for simple backgrounds, because the linear approximation neglects interactions of EM fields in complex media. In highly inhomogeneous media, nonlinear EM imaging is required to quantitatively evaluate the target. SDM provides a way that seamlessly combines ML and EM modeling, but the numerical forward modeling process limits its online prediction speed. Strategies unrolling PDE-based forward and inverse processes on neural networks for higher speed are proposed. 
	
	The methods in \cite{guo2021physics2} and  \cite{9539099} are applicable when the  background medium is known a priori (so that the Green's function is known). They originate from different views, i.e., the mathematical and physical views. The former mimics the optimization process, while the latter progressively retrieves the permittivity by predicting more and more accurate electric field, which is equivalent to taking more and more multiple scatterings into account. {\color{black}It is shown that partially unfolding forward and inverse problems according to the inverse scattering theory leads to a time- and memory-efficient neural network.}
	
	When the background medium is unknown, describing wave equations with partial differential operations is more appropriate.  The work in \cite{hu2021theory} shows the feasibility of embedding partial differential operations into the network. This direction has potentially many applications for reconstructing complex media, such as biomedicine engineering and geophysical exploration.
	
	{\color{black}
		\section{Challenges and opportunities }
		\label{sec:discussion}
		We tentatively discuss some open challenges and opportunities in this field from three aspects: data, physics, and algorithm.
		\vspace{-0.4cm}
		\subsection{Data}
		In detection, it is usually challenging to obtain the exact electric properties of targets, and the images in historical datasets often suffer from non-unique interpretations. Synthetically generating training data is time-consuming for large-scale imaging problems. Therefore, public datasets that cover various typical applications are required so that researchers can train, test, and compare among different methods. The datasets should contain scenarios where the inverse problem is highly ill-posed, e.g., highly inhomogeneous media, multiple strong scatterers, and limited observations, to test the performance limit of imaging algorithms. In addition, the methodology for evaluating the completeness of training datasets should be investigated \cite{7530821}.
		
		\vspace{-0.4cm}
		\subsection{Physics}
		Incorporating physics theory into data-driven methods is challenging. The \ac{doi} is partitioned into triangle (2D) or tetrahedral (3D) elements for accurate EM modeling in many applications. The element size, number, and topology are different case by case, which means that the number of neurons and their connections vary with training samples. Graph neural network may provide a solution \cite{9633179,al2021graph} to this challenge but  needs further investigation. For large-scale EM problems, the number of elements may be millions or billions. Training cost will be a critical issue when the elements are expressed as neurons. Finally, DL techniques also provide new perspectives of integrating EM methods with other imaging modalities to achieve better resolution \cite{sun2020deep,mojabi2021cnn,guo2020joint}, but how to embed different physical principles in a unified neural network remains open.
		
		\vspace{-0.4cm}
		\subsection{Algorithm}
		The credibility of predictions needs to be improved. Current strategies mainly rely on statistical analyses of test datasets. An alternative may be checking data fitness after image reconstruction. Other methods, such as uncertainty analysis, are in urgent need \cite{wei2020uncertainty,oh2021bayesian}. In addition, performance guarantees can be further investigated. The physics embedded DL structures increase interpretability, which may benefit the progress of theoretical guarantees and thus release the burden on parameter tuning. Furthermore, while  mean-squared error is mainly used as the optimization target in EM imaging, loss functions that minimize structure similarity \cite{huang2020deep}, image features \cite{zheng2022ct}, or probability distribution \cite{9096387} can be flexibly applied in the DL framework.}
	
	\section{Conclusions and outlooks}
	\label{sec:conclusion}
	
	We have surveyed three types of physics embedded data-driven imaging methods. \textit{Learning after physics processing} is straightforward, and it allows to borrow advanced DL techniques in image processing at a minimal cost. The shortcoming is that the DNN training does not follow wave physics, leading to difficulties in processing out-of-distribution data. {\color{black}Therefore, it is suitable for fast estimation of target shapes and properties but has risks in quantitative imaging. }

	\textit{Learning with physics loss} ensures training to obey wave physics and reduces the ill-posedness of imaging, {\color{black}so that the neural networks are more robust than \textit{learning after physics processing}}. The solution process for PDE and the inverse problem can be simultaneously trained using PINN; however, it requires a large volume of data that may be difficult to collect in practice. Using measurement loss function based on a predefined forward problem can perform better for inverting limited-aperture data.  One limitation of this method is that physics cannot guide image reconstruction in online prediction. {\color{black}Compared to \textit{learning after physics processing}, methods of this type require more computation resources in the training stage.}
	
	\textit{Learning with physics models} incorporates physics operators in both training and prediction. Imaging mainly relies on physics computation, while the trained parameters provide necessary prior knowledge for image reconstruction. {\color{black}It has better generalizability than former methods and can achieve real-time imaging. However, the neural network architecture needs to be tailored for different problems, increasing the complexity in the design.} 
	
	Artificial intelligence has blossomed in recent years due to the advancement of modern hardware and software, which builds the foundations of these progresses. While the EM community is glad to embrace these changes, the reliability of data-driven imaging remains an issue. EM theory provides baselines for EM sensing and imaging. Embedding physics in DNNs can improve interpretability and generalizability and thus improve safety in real-world applications. Recent research in EM imaging has proven its feasibility; we are sure this path will continue to expand in the next few years.

	
	\begin{spacing}{1.0}
		\bibliographystyle{ieeetr}
		\bibliography{IEEEexample_v2}

\begin{thebibliography}{10}

\bibitem{chew1990waves}
W.~Chew, {\em Waves and fields in inhomogeneous media}.
\newblock Springer, 1990.

\bibitem{massa2019dnns}
A.~Massa, D.~Marcantonio, X.~Chen, M.~Li, and M.~Salucci, ``{DNNs} as applied
  to electromagnetics, antennas, and propagation—a review,'' {\em IEEE
  Antennas and Wireless Propagation Letters}, vol.~18, no.~11, pp.~2225--2229,
  2019.

\bibitem{wang2021deep}
G.~Wang, M.~Jacob, X.~Mou, Y.~Shi, and Y.~C. Eldar, ``Deep tomographic image
  reconstruction: Yesterday, today, and tomorrow—editorial for the 2nd
  special issue “machine learning for image reconstruction”,'' {\em IEEE
  Transactions on Medical Imaging}, vol.~40, no.~11, pp.~2956--2964, 2021.

\bibitem{li2021machine}
M.~Li, R.~Guo, K.~Zhang, Z.~Lin, F.~Yang, S.~Xu, X.~Chen, A.~Massa, and
  A.~Abubakar, ``Machine learning in electromagnetics with applications to
  biomedical imaging: A review,'' {\em IEEE Antennas and Propagation Magazine},
  vol.~63, no.~3, pp.~39--51, 2021.

\bibitem{chen2020review}
X.~Chen, Z.~Wei, M.~Li, and P.~Rocca, ``A review of deep learning approaches
  for inverse scattering problems (invited review),'' {\em Progress In
  Electromagnetics Research}, vol.~167, pp.~67--81, 2020.

\bibitem{sanghvi2019embedding}
Y.~Sanghvi, Y.~Kalepu, and U.~K. Khankhoje, ``Embedding deep learning in
  inverse scattering problems,'' {\em IEEE Transactions on Computational
  Imaging}, vol.~6, pp.~46--56, 2019.

\bibitem{lin2021low}
Z.~Lin, R.~Guo, M.~Li, A.~Abubakar, T.~Zhao, F.~Yang, and S.~Xu,
  ``Low-frequency data prediction with iterative learning for highly nonlinear
  inverse scattering problems,'' {\em IEEE Transactions on Microwave Theory and
  Techniques}, vol.~69, no.~10, pp.~4366--4376, 2021.

\bibitem{bora2017compressed}
A.~Bora, A.~Jalal, E.~Price, and A.~G. Dimakis, ``Compressed sensing using
  generative models,'' in {\em International Conference on Machine Learning},
  pp.~537--546, PMLR, 2017.

\bibitem{8565987}
L.~Li, L.~G. Wang, F.~L. Teixeira, C.~Liu, A.~Nehorai, and T.~J. Cui,
  ``{DeepNIS}: Deep neural network for nonlinear electromagnetic inverse
  scattering,'' {\em IEEE Transactions on Antennas and Propagation}, vol.~67,
  no.~3, pp.~1819--1825, 2019.

\bibitem{wei2018deep}
Z.~Wei and X.~Chen, ``Deep-learning schemes for full-wave nonlinear inverse
  scattering problems,'' {\em IEEE Transactions on Geoscience and Remote
  Sensing}, vol.~57, no.~4, pp.~1849--1860, 2018.

\bibitem{ye2020inhomogeneous}
X.~Ye, Y.~Bai, R.~Song, K.~Xu, and J.~An, ``An inhomogeneous background imaging
  method based on generative adversarial network,'' {\em IEEE Transactions on
  Microwave Theory and Techniques}, vol.~68, no.~11, pp.~4684--4693, 2020.

\bibitem{9751403}
X.~Ye, D.~Yang, X.~Yuan, R.~Song, S.~Sun, and D.~Fang, ``Application of
  generative adversarial network-based inversion algorithm in imaging
  two-dimensional lossy biaxial anisotropic scatterer,'' {\em IEEE Transactions
  on Antennas and Propagation}, pp.~1--1, 2022.

\bibitem{jin2020physics}
Y.~Jin, Q.~Shen, X.~Wu, J.~Chen, and Y.~Huang, ``A physics-driven deep-learning
  network for solving nonlinear inverse problems,'' {\em Petrophysics-The SPWLA
  Journal of Formation Evaluation and Reservoir Description}, vol.~61, no.~01,
  pp.~86--98, 2020.

\bibitem{raissi2019physics}
M.~Raissi, P.~Perdikaris, and G.~E. Karniadakis, ``Physics-informed neural
  networks: A deep learning framework for solving forward and inverse problems
  involving nonlinear partial differential equations,'' {\em Journal of
  Computational Physics}, vol.~378, pp.~686--707, 2019.

\bibitem{bar2021strong}
L.~Bar and N.~Sochen, ``Strong solutions for {PDE}-based tomography by
  unsupervised learning,'' {\em SIAM Journal on Imaging Sciences}, vol.~14,
  no.~1, pp.~128--155, 2021.

\bibitem{9539099}
J.~Liu, H.~Zhou, T.~Ouyang, Q.~Liu, and Y.~Wang, ``Physical model-inspired deep
  unrolling network for solving nonlinear inverse scattering problems,'' {\em
  IEEE Transactions on Antennas and Propagation}, vol.~70, no.~2,
  pp.~1236--1249, 2022.

\bibitem{8434321}
H.~K. Aggarwal, M.~P. Mani, and M.~Jacob, ``{MoDL}: Model-based deep learning
  architecture for inverse problems,'' {\em IEEE Transactions on Medical
  Imaging}, vol.~38, no.~2, pp.~394--405, 2019.

\bibitem{guo2021physics1}
R.~Guo, Z.~Lin, T.~Shan, X.~Song, M.~Li, F.~Yang, S.~Xu, and A.~Abubakar,
  ``Physics embedded deep neural network for solving full-wave inverse
  scattering problems,'' {\em IEEE Transactions on Antennas and Propagation},
  2021.

\bibitem{shahriari2021error}
M.~Shahriari, D.~Pardo, J.~A. Rivera, C.~Torres-Verd{\'\i}n, A.~Picon,
  J.~Del~Ser, S.~Ossand{\'o}n, and V.~M. Calo, ``Error control and loss
  functions for the deep learning inversion of borehole resistivity
  measurements,'' {\em International Journal for Numerical Methods in
  Engineering}, vol.~122, no.~6, pp.~1629--1657, 2021.

\bibitem{hu2021theory}
Y.~Hu, Y.~Jin, X.~Wu, and J.~Chen, ``A theory-guided deep neural network for
  time domain electromagnetic simulation and inversion using a differentiable
  programming platform,'' {\em IEEE Transactions on Antennas and Propagation},
  2021.

\bibitem{guo2020pixel}
R.~Guo, Z.~Jia, X.~Song, M.~Li, F.~Yang, S.~Xu, and A.~Abubakar, ``Pixel-and
  model-based microwave inversion with supervised descent method for dielectric
  targets,'' {\em IEEE Transactions on Antennas and Propagation}, vol.~68,
  no.~12, pp.~8114--8126, 2020.

\bibitem{guo2021physics2}
R.~Guo, T.~Shan, X.~Song, M.~Li, F.~Yang, S.~Xu, and A.~Abubakar, ``Physics
  embedded deep neural network for solving volume integral equation: 2d case,''
  {\em IEEE Transactions on Antennas and Propagation}, 2021.

\bibitem{Fu2021Toep}
R.~Fu, Y.~Liu, T.~Huang, and Y.~C. Eldar, ``Structured {LISTA} for
  multidimensional harmonic retrieval,'' {\em IEEE Transactions on Signal
  Processing}, vol.~69, pp.~3459--3472, 2021.

\bibitem{fu2022block}
R.~Fu, T.~Huang, L.~Wang, and Y.~Liu, ``Block-sparse recovery network for
  two-dimensional harmonic retrieval,'' {\em Electronics Letters}, vol.~58,
  no.~6, pp.~249--251, 2022.

\bibitem{shan2021neural}
T.~Shan, Z.~Lin, X.~Song, M.~Li, F.~Yang, and S.~Xu, ``Neural born iteration
  method for solving inverse scattering problems: {2D} cases,'' {\em arXiv
  preprint arXiv:2112.09831}, 2021.

\bibitem{shlezinger2020model}
N.~Shlezinger, J.~Whang, Y.~C. Eldar, and A.~G. Dimakis, ``Model-based deep
  learning,'' {\em arXiv preprint arXiv:2012.08405}, 2020.

\bibitem{habashy2004general}
T.~Habashy and A.~Abubakar, ``A general framework for constraint minimization
  for the inversion of electromagnetic measurements,'' {\em Progress in
  Electromagnetics search}, vol.~46, pp.~265--312, 2004.

\bibitem{jin2011theory}
J.-M. Jin, {\em Theory and computation of electromagnetic fields}.
\newblock John Wiley \& Sons, 2011.

\bibitem{klose2022laterally}
T.~Klose, J.~Guillemoteau, G.~Vignoli, and J.~Tronicke, ``Laterally constrained
  inversion ({LCI}) of multi-configuration {EMI} data with tunable sharpness,''
  {\em Journal of Applied Geophysics}, vol.~196, p.~104519, 2022.

\bibitem{zhdanov2009new}
M.~S. Zhdanov, ``New advances in regularized inversion of gravity and
  electromagnetic data,'' {\em Geophysical Prospecting}, vol.~57, no.~4,
  pp.~463--478, 2009.

\bibitem{vignoli2015sharp}
G.~Vignoli, G.~Fiandaca, A.~V. Christiansen, C.~Kirkegaard, and E.~Auken,
  ``Sharp spatially constrained inversion with applications to transient
  electromagnetic data,'' {\em Geophysical Prospecting}, vol.~63, no.~1,
  pp.~243--255, 2015.

\bibitem{abubakar2002imaging}
A.~Abubakar, P.~M. Van~den Berg, and J.~J. Mallorqui, ``Imaging of biomedical
  data using a multiplicative regularized contrast source inversion method,''
  {\em IEEE Transactions on Microwave Theory and Techniques}, vol.~50, no.~7,
  pp.~1761--1771, 2002.

\bibitem{zhong2021electrical}
S.~Zhong, Y.~Wang, Y.~Zheng, S.~Wu, X.~Chang, and W.~Zhu, ``Electrical
  resistivity tomography with smooth sparse regularization,'' {\em Geophysical
  Prospecting}, vol.~69, no.~8-9, pp.~1773--1789, 2021.

\bibitem{vignoli2005focusing}
G.~Vignoli and L.~Zanzi, ``Focusing inversion technique applied to radar
  tomographic data,'' in {\em Near surface 2005-11th European meeting of
  environmental and engineering geophysics}, pp.~cp--13, European Association
  of Geoscientists \& Engineers, 2005.

\bibitem{Potter2009_radarlp}
L.~C. Potter, E.~Ertin, J.~T. Parker, and M.~Cetin, ``Sparsity and compressed
  sensing in radar imaging,'' {\em Proceedings of the IEEE}, vol.~98, no.~6,
  pp.~1006--1020, 2010.

\bibitem{Patel2009_SARl1}
V.~M. Patel, G.~R. Easley, D.~M. Healy, and R.~Chellappa, ``Compressed
  synthetic aperture radar,'' {\em IEEE Journal of Selected Topics in Signal
  Processing}, vol.~4, no.~2, pp.~244--254, 2010.

\bibitem{zhdanov20003d}
M.~Zhdanov and G.~Hursan, ``{3D} electromagnetic inversion based on
  quasi-analytical approximation,'' {\em Inverse Problems}, vol.~16, no.~5,
  p.~1297, 2000.

\bibitem{christiansen2016efficient}
A.~V. Christiansen, E.~Auken, C.~Kirkegaard, C.~Schamper, and G.~Vignoli, ``An
  efficient hybrid scheme for fast and accurate inversion of airborne transient
  electromagnetic data,'' {\em Exploration Geophysics}, vol.~47, no.~4,
  pp.~323--330, 2016.

\bibitem{shen2018data}
Q.~Shen, J.~Chen, and H.~Wang, ``Data-driven interpretation of ultradeep
  azimuthal propagation resistivity measurements: Transdimensional stochastic
  inversion and uncertainty quantification,'' {\em Petrophysics-The SPWLA
  Journal of Formation Evaluation and Reservoir Description}, vol.~59, no.~06,
  pp.~786--798, 2018.

\bibitem{hansen2021efficient}
T.~M. Hansen, ``Efficient probabilistic inversion using the rejection
  sampler—exemplified on airborne {EM} data,'' {\em Geophysical Journal
  International}, vol.~224, no.~1, pp.~543--557, 2021.

\bibitem{de2019probabilistic}
G.~De~Pasquale, N.~Linde, J.~Doetsch, and W.~S. Holbrook, ``Probabilistic
  inference of subsurface heterogeneity and interface geometry using
  geophysical data,'' {\em Geophysical Journal International}, vol.~217, no.~2,
  pp.~816--831, 2019.

\bibitem{xu2020deep}
K.~Xu, L.~Wu, X.~Ye, and X.~Chen, ``Deep learning-based inversion methods for
  solving inverse scattering problems with phaseless data,'' {\em IEEE
  Transactions on Antennas and Propagation}, vol.~68, no.~11, pp.~7457--7470,
  2020.

\bibitem{8476623}
Z.~Wei and X.~Chen, ``Deep-learning schemes for full-wave nonlinear inverse
  scattering problems,'' {\em IEEE Transactions on Geoscience and Remote
  Sensing}, vol.~57, no.~4, pp.~1849--1860, 2019.

\bibitem{xiao2019fast}
J.~Xiao, J.~Li, Y.~Chen, F.~Han, and Q.~H. Liu, ``Fast electromagnetic
  inversion of inhomogeneous scatterers embedded in layered media by {Born}
  approximation and 3-{D} {U}-{N}et,'' {\em IEEE Geoscience and Remote Sensing
  Letters}, vol.~17, no.~10, pp.~1677--1681, 2019.

\bibitem{wei2019physics}
Z.~Wei and X.~Chen, ``Physics-inspired convolutional neural network for solving
  full-wave inverse scattering problems,'' {\em IEEE Transactions on Antennas
  and Propagation}, vol.~67, no.~9, pp.~6138--6148, 2019.

\bibitem{wei2020uncertainty}
Z.~Wei and X.~Chen, ``Uncertainty quantification in inverse scattering problems
  with {B}ayesian convolutional neural networks,'' {\em IEEE Transactions on
  Antennas and Propagation}, vol.~69, no.~6, pp.~3409--3418, 2020.

\bibitem{abubakar20082}
A.~Abubakar, T.~Habashy, V.~Druskin, L.~Knizhnerman, and D.~Alumbaugh, ``2.5
  {D} forward and inverse modeling for interpreting low-frequency
  electromagnetic measurements,'' {\em Geophysics}, vol.~73, no.~4,
  pp.~F165--F177, 2008.

\bibitem{shahriari2020deep}
M.~Shahriari, D.~Pardo, A.~Pic{\'o}n, A.~Galdran, J.~Del~Ser, and
  C.~Torres-Verd{\'\i}n, ``A deep learning approach to the inversion of
  borehole resistivity measurements,'' {\em Computational Geosciences},
  vol.~24, no.~3, pp.~971--994, 2020.

\bibitem{karniadakis2021physics}
G.~E. Karniadakis, I.~G. Kevrekidis, L.~Lu, P.~Perdikaris, S.~Wang, and
  L.~Yang, ``Physics-informed machine learning,'' {\em Nature Reviews Physics},
  vol.~3, no.~6, pp.~422--440, 2021.

\bibitem{sahel2022deep}
Y.~B. Sahel, J.~P. Bryan, B.~Cleary, S.~L. Farhi, and Y.~C. Eldar, ``Deep
  unrolled recovery in sparse biological imaging: Achieving fast, accurate
  results,'' {\em IEEE Signal Processing Magazine}, vol.~39, no.~2, pp.~45--57,
  2022.

\bibitem{monga2021algorithm}
V.~Monga, Y.~Li, and Y.~C. Eldar, ``Algorithm unrolling: Interpretable,
  efficient deep learning for signal and image processing,'' {\em IEEE Signal
  Processing Magazine}, vol.~38, no.~2, pp.~18--44, 2021.

\bibitem{Beck2009A}
A.~Beck and M.~Teboulle, ``A fast iterative shrinkage-thresholding algorithm
  for linear inverse problems,'' {\em Siam J Imaging Sciences}, vol.~2, no.~1,
  pp.~183--202, 2009.

\bibitem{Gregor2010Learning}
K.~Gregor and Y.~Lecun, ``Learning fast approximations of sparse coding,'' in
  {\em International Conference on International Conference on Machine
  Learning}, pp.~399--406, 2010.

\bibitem{xiong2013supervised}
X.~Xiong and F.~De~la Torre, ``Supervised descent method and its applications
  to face alignment,'' in {\em Proceedings of the IEEE Conference on Computer
  Vision and Pattern Recognition}, pp.~532--539, 2013.

\bibitem{guo2020application}
R.~Guo, M.~Li, F.~Yang, S.~Xu, and A.~Abubakar, ``Application of supervised
  descent method for {2D} magnetotelluric data inversion,'' {\em Geophysics},
  vol.~85, no.~4, pp.~WA53--WA65, 2020.

\bibitem{hu2020supervised}
Y.~Hu, R.~Guo, Y.~Jin, X.~Wu, M.~Li, A.~Abubakar, and J.~Chen, ``A supervised
  descent learning technique for solving directional electromagnetic
  logging-while-drilling inverse problems,'' {\em IEEE Transactions on
  Geoscience and Remote Sensing}, vol.~58, no.~11, pp.~8013--8025, 2020.

\bibitem{lu20211}
S.~Lu, B.~Liang, J.~Wang, F.~Han, and Q.~H. Liu, ``1-{D} inversion of {GREATEM}
  data by supervised descent learning,'' {\em IEEE Geoscience and Remote
  Sensing Letters}, vol.~19, pp.~1--5, 2022.

\bibitem{hao2021robust}
P.~Hao, X.~Sun, Z.~Nie, X.~Yue, and Y.~Zhao, ``A robust inversion of induction
  logging responses in anisotropic formation based on supervised descent
  method,'' {\em IEEE Geoscience and Remote Sensing Letters}, vol.~19,
  pp.~1--5, 2022.

\bibitem{jia20213}
Z.~Jia, R.~Guo, M.~Li, G.~Wang, Z.~Liu, and Y.~Shao, ``3-{D} model-based
  inversion using supervised descent method for aspect-limited microwave data
  of metallic targets,'' {\em IEEE Transactions on Geoscience and Remote
  Sensing}, 2021.

\bibitem{lin2020neural}
Z.~Lin, R.~Guo, K.~Zhang, M.~Li, F.~Yang, S.~Xu, and A.~Abubakar, ``Neural
  network-based supervised descent method for {2D} electrical impedance
  tomography,'' {\em Physiological Measurement}, vol.~41, no.~7, p.~074003,
  2020.

\bibitem{zhang2020supervised}
K.~Zhang, R.~Guo, M.~Li, F.~Yang, S.~Xu, and A.~Abubakar, ``Supervised descent
  learning for thoracic electrical impedance tomography,'' {\em IEEE
  Transactions on Biomedical Engineering}, vol.~68, no.~4, pp.~1360--1369,
  2020.

\bibitem{guo2019regularized}
R.~Guo, M.~Li, F.~Yang, S.~Xu, and A.~Abubakar, ``Regularized supervised
  descent method for 2-{D} magnetotelluric data inversion,'' in {\em SEG
  Technical Program Expanded Abstracts 2019}, pp.~2508--2512, Society of
  Exploration Geophysicists, 2019.

\bibitem{guo2021electromagnetic}
L.~Guo, M.~Li, S.~Xu, F.~Yang, and L.~Liu, ``Electromagnetic modeling using an
  {FDTD}-equivalent recurrent convolution neural network: Accurate computing on
  a deep learning framework.,'' {\em IEEE Antennas and Propagation Magazine},
  2021.

\bibitem{van1997contrast}
P.~M. Van Den~Berg and R.~E. Kleinman, ``A contrast source inversion method,''
  {\em Inverse Problems}, vol.~13, no.~6, p.~1607, 1997.

\bibitem{7530821}
M.~Salucci, N.~Anselmi, G.~Oliveri, P.~Calmon, R.~Miorelli, C.~Reboud, and
  A.~Massa, ``Real-time {NDT}-{NDE} through an innovative adaptive partial
  least squares {SVR} inversion approach,'' {\em IEEE Transactions on
  Geoscience and Remote Sensing}, vol.~54, no.~11, pp.~6818--6832, 2016.

\bibitem{9633179}
W.~Herzberg, D.~B. Rowe, A.~Hauptmann, and S.~J. Hamilton, ``Graph
  convolutional networks for model-based learning in nonlinear inverse
  problems,'' {\em IEEE Transactions on Computational Imaging}, vol.~7,
  pp.~1341--1353, 2021.

\bibitem{al2021graph}
A.~Al-Saffar, L.~Guo, and A.~Abbosh, ``Graph attention network for microwave
  imaging of brain anomaly,'' {\em arXiv preprint arXiv:2108.01965}, 2021.

\bibitem{sun2020deep}
Y.~Sun, B.~Denel, N.~Daril, L.~Evano, P.~Williamson, and M.~Araya-Polo, ``Deep
  learning joint inversion of seismic and electromagnetic data for salt
  reconstruction,'' in {\em SEG Technical Program Expanded Abstracts 2020},
  pp.~550--554, Society of Exploration Geophysicists, 2020.

\bibitem{mojabi2021cnn}
P.~Mojabi, M.~Hughson, V.~Khoshdel, I.~Jeffrey, and J.~LoVetri, ``{CNN} for
  compressibility to permittivity mapping for combined ultrasound-microwave
  breast imaging,'' {\em IEEE Journal on Multiscale and Multiphysics
  Computational Techniques}, vol.~6, pp.~62--72, 2021.

\bibitem{guo2020joint}
R.~Guo, H.~M. Yao, M.~Li, M.~K.~P. Ng, L.~Jiang, and A.~Abubakar, ``Joint
  inversion of audio-magnetotelluric and seismic travel time data with deep
  learning constraint,'' {\em IEEE Transactions on Geoscience and Remote
  Sensing}, vol.~59, no.~9, pp.~7982--7995, 2021.

\bibitem{oh2021bayesian}
S.~Oh and J.~Byun, ``Bayesian uncertainty estimation for deep learning
  inversion of electromagnetic data,'' {\em IEEE Geoscience and Remote Sensing
  Letters}, vol.~19, pp.~1--5, 2022.

\bibitem{huang2020deep}
Y.~Huang, R.~Song, K.~Xu, X.~Ye, C.~Li, and X.~Chen, ``Deep learning-based
  inverse scattering with structural similarity loss functions,'' {\em IEEE
  Sensors Journal}, vol.~21, no.~4, pp.~4900--4907, 2020.

\bibitem{zheng2022ct}
A.~Zheng, K.~Liang, L.~Zhang, and Y.~Xing, ``A {CT} image feature space
  ({CTIS}) loss for restoration with deep learning-based methods,'' {\em
  Physics in Medicine \& Biology}, vol.~67, no.~5, p.~055010, 2022.

\bibitem{9096387}
Z.~Hu, H.~Xue, Q.~Zhang, J.~Gao, N.~Zhang, S.~Zou, Y.~Teng, X.~Liu, Y.~Yang,
  D.~Liang, X.~Zhu, and H.~Zheng, ``{DPIR-Net}: Direct {PET} image
  reconstruction based on the {Wasserstein} generative adversarial network,''
  {\em IEEE Transactions on Radiation and Plasma Medical Sciences}, vol.~5,
  no.~1, pp.~35--43, 2021.

\end{thebibliography}
	\end{spacing} 
	
	
\end{document}